\documentclass[%
nofootinbib, 
 amsmath,amssymb,
 aps,
 prd,
twocolumn, superscriptaddress
]{revtex4-1}

\usepackage[normalem]{ulem}

\usepackage{graphicx}
\usepackage{dcolumn}
\usepackage{bm}
\usepackage{hyperref}
\usepackage{amssymb}
\usepackage{xcolor}
\usepackage[dvipsnames]{xcolor}
\hypersetup{colorlinks,citecolor=OrangeRed}
\hypersetup{colorlinks=true, linkcolor=blue, urlcolor=MidnightBlue}

\usepackage{orcidlink}

\begin{document}

\preprint{APS/123-QED}

\title{Dark energy stars from the modified Chaplygin gas: $C-I-\Lambda-E_g-f$ universal relations}

\author{Krishna Pada Das \orcidlink{0000-0003-4653-6006}}
 \email{krishnapada0019@gmail.com}
 \affiliation{Department of Mathematics, Indian Institute of Engineering Science and Technology, \\ Shibpur, Howrah, 711 103, India
}

\author{Juan M. Z. Pretel \orcidlink{0000-0003-0883-3851}}
 \email{juan04manuel91@gmail.com}
 \affiliation{Centro Brasileiro de Pesquisas F{\'i}sicas, Rua Dr.~Xavier Sigaud, 150 URCA, Rio de Janeiro CEP 22290-180, RJ, Brazil
}

\date{\today}

\begin{abstract}
Dark energy stars (DESs), described by the modified Chaplygin gas (MCG), can be dynamically stable and fall within different observational measurements. In this work, we employ diverse macroscopic properties, such as compactness $C$, moment of inertia $I$, tidal deformability $\Lambda$, gravitational binding energy $E_g$ and $f$-mode nonradial pulsation frequency, to explore whether they are correlated by universal relations (URs). Remarkably, our stellar configurations always obey the causality condition and are compatible with several observational mass-radius constraints. Via the $C-I-\text{Love}-f$ URs, our results reveal that we cannot distinguish quark stars (QSs) from DESs in the sense that DESs satisfy several URs very similar to those of QSs. However, when we involve $E_g$, DESs and QSs can be strongly distinguished through the $I-E_g^{-2}$, $\Lambda-E_g^{-5}$ and $f-E_g^{-2}$ URs. We also make use of these findings and the tidal deformability constraint from the GW170817 event to forecast the canonical properties of a $1.4\, M_\odot$ compact star. Furthermore, we present a set of fine empirical correlations involving the tidal deformability, obtained from an extensive scan of the parameter space of our DE stellar models.
\end{abstract}

\maketitle


\section{Introduction}
After Hubble established cosmic expansion in 1929, it was widely believed that gravity would gradually slow such expansion. Nevertheless, late-20th-century observations of distant Type Ia supernovae revealed that the expansion of the universe is accelerating. This unexpected result is one of the most remarkable discoveries in modern cosmology, first reported in studies \cite{perlmutter1998discovery, riess1998observational, hicken2009improved, seikel2009model}. Although the scientific community did not anticipate this finding, it led to the widely accepted idea that an unknown component pervades the universe and drives the observed acceleration. This mysterious fluid component must possess certain characteristics: it must exert negative pressure, be homogeneous, be present everywhere in space, and not interact with the fundamental forces, except gravity. Despite extensive research, the true origin and nature of this mysterious component remain unknown. To account for this phenomenon, the term dark energy (DE) has been introduced to describe its elusive nature. Based on multiple observations, the present-day energy budget of the universe is composed of approximately 5\% ordinary baryonic matter and radiation, 26–27\% dark matter (DM), and 68–70\% DE \cite{aghanim2020planck}.

To address these challenges, numerous models have been proposed to explain the origin and nature of DE, including a wide variety of modified gravity theories. DE is typically characterized as a form of exotic fluid with negative pressure that violates the strong energy condition, expressed mathematically as $\rho + 3p < 0$, where $\rho$ is the energy density and $p$ is the pressure. In general, the equation of state (EoS) for DE is given by $p=w\rho$ providing a useful classification of DE models. For values in the range $-1< w <-1/3$, DE is commonly referred to as \textit{quintessence}. The special case $w = -1$ corresponds to a cosmological constant, while models with $w<-1$ are known as \textit{phantom} or \textit{ghost-like} DE and typically violate the null energy condition. However, the literature also provides other negative pressure phenomenological EoSs to describe cosmic acceleration \cite{Kamenshchik2001}.

The inhomogeneous Chaplygin gas (CG) has been studied as a simple unified model for DM and DE, motivated by a geometric framework inspired by M-theory \cite{bilic2002unification}. It has also been suggested that if the universe is dominated by the CG, the existence of a cosmological constant can be confidently excluded \cite{Makler2003}. Regarding DE, we adopt the modified Chaplygin gas (MCG) EoS \cite{debnath2004role, pourhassan2014extended}, originally proposed as a unified version for DE and non-relativistic matter, treating them as a single fluid rather than two distinct components. 
The MCG is characterized by an EoS of the form
\begin{equation}\label{EqoS}
    p = A\rho - \frac{B}{\rho^\alpha} , 
\end{equation}
which is an example of a DE model and has three degrees of freedom $\{A,B,\alpha\}$, where $A$ is a positive dimensionless constant, $B$ is a positive parameter given in $\rm m^{-2(1+\alpha)}$ units in a geometric unit system, and the extra constant $\alpha$ satisfies $0 \leq \alpha \leq 1$ \cite{Zheng2022}. In particular, when $A=0$, it corresponds to an exotic background fluid with negative pressure and leads to the accelerated
expansion of the universe, i.e.,~the original CG model, described by the EoS $p= -B/\rho$ for $\alpha= 1$ \cite{Kamenshchik2001}, while for $\alpha \neq 1$ we recover the generalized CG model \cite{Bento2003, Lu2022, Lian2021}, given by $p= -B/\rho^\alpha$. When $B=0$, it reduces to the traditional EoS of a perfect fluid; $p= A\rho$. Furthermore, the MCG model reduces to the standard $\Lambda$CDM model when $A=\alpha= 0$. Consequently, our work will employ the most general version of the Chaplygin gas to describe DESs, thus generalizing previous studies.

In the framework of the MCG model, the negative-pressure term $-B/\rho^{\alpha}$ is negligible at high densities and becomes significant only in the low-density regime. Accordingly, during the early stages of collapse the positive barotropic term $A\rho$ dominates, and the fluid behaves like ordinary matter undergoing gravitational accumulation. As the system evolves and the effective density decreases (or a critical surface is approached), the negative-pressure component becomes dominant, producing a repulsive effect that halts further collapse and prevents singularity formation, leading instead to a DE-dominated core. This behavior is well established in the literature on Chaplygin-type EoSs \cite{bento2002generalized,debnath2004role}. The parameter $\alpha$ serves as a phenomenological index that determines how the negative-pressure component varies with the energy density. In effect, it governs the stiffness of the fluid and the pressure-density relationship at low densities, as well as the rate at which the model shifts from a dust-dominated regime to a dark-energy-dominated regime. Smaller values of $\alpha$ produce a smoother, more gradual transition, while larger values produce a sharper change between these two behaviors.

\begin{figure}
    \centering
    \includegraphics[width=8.3cm]{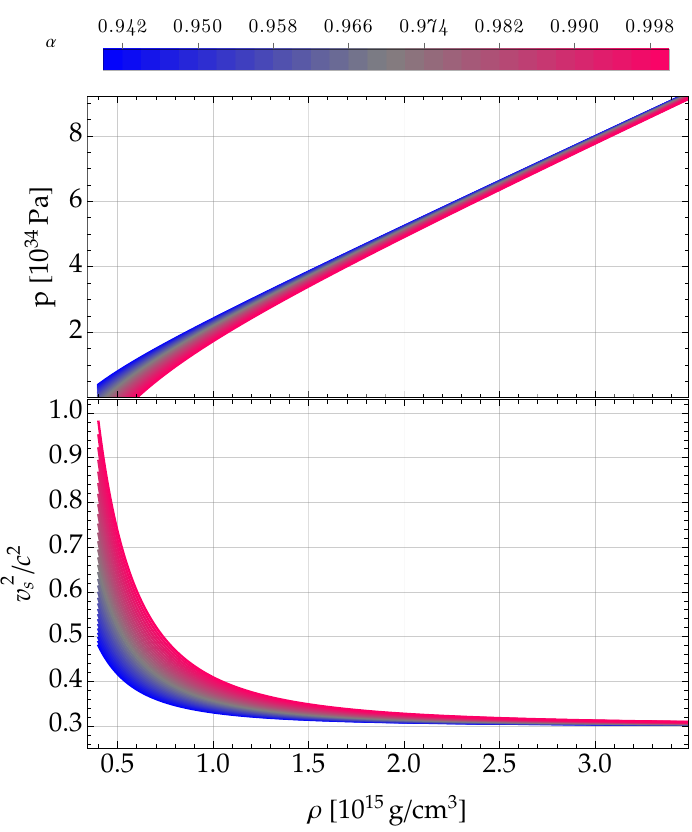}
    \caption{Pressure \eqref{EqoS} (top panel) and squared speed of sound \eqref{SpeedofVeloEq} (lower panel) as a function of mass density, by adopting the MCG parameters $A= 0.3$, $B= 6.0\times 10^{-20}\, \rm m^{-2(1+\alpha)}$ and varying $\alpha$ in the range $\alpha \in [0.94, 1.0]$. The higher the $\alpha$, the lower the pressure, but there is an increase in $v_s^2$. }
    \label{FigEoSandSpeedSound}
\end{figure}

The speed of sound inside compact stars, such as neutron stars (NSs) and quark stars (QSs), is a topic of significant interest in astrophysics because it provides information on how sound waves propagate in the stellar medium. Additionally, as an important requirement, the speed of sound (defined by $v_s= \sqrt{dp/d\rho}$) must be less than the speed of light to respect the causality condition. Consequently, in view of the EoS \eqref{EqoS}, the squared speed of sound for DESs with a Chaplygin dark fluid (CDF) is given by
\begin{equation}\label{SpeedofVeloEq}
    v_s^2 = \frac{dp}{d\rho} = A+ \frac{\alpha B}{\rho^{1+\alpha}} ,
\end{equation}
so that, for realistic fluids one requires $0\leq v_s^{2} \leq 1$ (positive for stability, and less than unity for causality). This condition is automatically satisfied for many cosmological densities only if $0\leq \alpha \leq 1$. For $\alpha=0$, the fluid behaves like $p = A\rho - B$, i.e., a cosmological constant plus a barotropic fluid. For $\alpha=1$, one recovers the standard Chaplygin gas. Values of $\alpha$ outside the interval $[0,1]$ tend to produce superluminal sound speeds, negative compressibility, or other unphysical early–universe behaviours.

The impact of varying the parameters $A$ and $B$ with fixed $\alpha= 1$ on the global properties of DESs has been widely investigated in recent years \cite{panotopoulos2021slowly, pretel2023radial, Bhattacharjee2024, Jyothilakshmi2024,banerjee2025EPJC}. We should mention that Panotopoulos recently found the $C-I-\Lambda$ correlations for Chaplygin fluid spheres considering the special case $\alpha= 1$ and three different models varying the constant parameters $A$ and $B$ \cite{panotopoulos2025UR}. Specifically, it was shown that variations in the Chaplygin-type EoS introduced by the three models lead to the same fit function for each UR. Therefore, in the present study we focus primarily on investigating the effects of $\alpha$ while keeping both $A$ and $B$ fixed as a main scenario (although in the Appendix \ref{extended} we will consider two more cases), and extend previous analyses by incorporating the gravitational binding energy and the $f$-mode nonradial oscillation frequency. We emphasize that, using our URs together with the tidal deformability constraint from GW170817, we derive theoretical bounds on the canonical properties of a $1.4\,M_\odot$ compact star, which are not directly measurable through astrophysical observations. With this in mind, we consider $A= 0.3$ and $B= 6.0\times 10^{-20}\, \rm m^{-2(1+\alpha)}$ and vary the MCG parameter $\alpha$ in the range $\alpha \in [0.94, 1.0]$. For these values, Fig.~\ref{FigEoSandSpeedSound} exhibits the behavior of the pressure and squared speed of sound as functions of the mass density. As expected, a small $\alpha$ increases the pressure but decreases $v_s^2$. Although causality is always obeyed for the range of densities, the total pressure $p$ can be negative above a certain value of $\alpha$ for small $\rho$. This means that our DES sequences have to be built taking care with the range of central densities so that the pressure is always positive. For example, for $\alpha=1$, the central density will be varied in the range $\rho\in [0.61,3.5]\times 10^{15}\, \rm g/cm^3$. We will see this later when determining the global properties of DESs but we affirm from now on that the causality condition is always respected in our DE stellar models, thus avoiding superluminal signal propagation.

If DE is indeed pervasive throughout the universe, a natural question arises as to whether it could also be present within astrophysical structures, such as hypothetical compact stars. This possibility suggests that DE may interact with ordinary matter in local astrophysical environments, including wormholes, black holes, compact stars, and other exotic compact objects. In addition, the study of alternative black hole models, which seek to resolve central singularities and modify the classical notion of event horizons, has become an increasingly active area of research in modern astroparticle physics. In this regard, various hypothetical compact objects have been proposed, including non-singular black holes \cite{dymnikova1992vacuum}, false vacuum bubbles \cite{coleman1980gravitational}, and gravastars \cite{mazur2004gravitational, mazur2023gravitational}, many of which are based on incorporating DE into stellar models. The earliest black hole solution assuming an EoS $p = w \rho$ with $w < -1/3$ was presented in Refs.~\cite{Kiselev2003,Fernando2012} and is referred to as a quintessential black hole. In addition, wormholes sustained by DE were investigated in Ref.~\cite{Sushkov2005Wormholes}, where the authors considered $\omega < -1$ to model phantom DE. Mazur and Mottola \cite{mazur2004gravitational, mazur2023gravitational} proposed hypothetical objects consisting of a stiff shell, whose surface layer is composed of stiff matter with positive pressure, connecting a DE core to the exterior vacuum through junction layers.

Drawing on these ideas, Chapline introduced the concept of DES \cite{chapline2005dark}, a hypothetical astrophysical object that contains a DE core and has a microscopic quantum critical layer instead of an event horizon. The fundamental principle underlying this model is that, as matter approaches a critical surface, it undergoes a transition into vacuum energy at a density substantially higher than the cosmological vacuum energy density. This conversion generates a strong negative pressure that counterbalances gravity \cite{chapline2005dark}, effectively preventing the formation of singularities. Given its near homogeneity and low density, the DE EoS provides an effective description of the internal fluid properties of compact stellar objects. In Ref.~\cite{Beltracchi2019}, a class of models for the formation of DESs was proposed, in which gravitational collapse evolves from an initial state with positive pressure to a final configuration containing a DE core defined by $p=-\rho$. These models, known as \emph{pileup models}, describe the gradual formation of the DE core through the accumulation of matter on its surface. During this process, the central energy density increases until it reaches a terminal value, while the pressure initially increases and then decreases to satisfy $p=-\rho$, signaling the onset of the DE core. Subsequently, the DE core expands outward.

In this work, we consider a hypothetical DES composed of MCG, and its possible formation mechanism is outlined below: From the MCG EoS (\ref{EqoS}), the term $A\rho$ corresponds to a positive barotropic pressure, while the term $-B/\rho^{\alpha}$ provides a negative-pressure component, analogous to a cosmological constant. At high densities, the $A\rho$ term dominates, leading to matter-like behavior and allowing gravitational collapse. As the density increases during collapse, the negative-pressure term becomes significant, introducing a repulsive force that opposes further compression. At sufficiently low densities, this term completely dominates and the MCG effectively behaves as a DE-like fluid. The resulting repulsive pressure can balance gravity, halting collapse and allowing the system to settle into a static equilibrium configuration. This equilibrium corresponds to a DE-dominated core, forming a compact, horizonless object known as a DES, supported by negative pressure rather than degeneracy or thermal effects.

Within the context of DES models, various investigations have addressed the characterization of these compact objects. In fact, Yazadjiev \cite{yazadjiev2011exact} examined the interior solution of compact stars composed of varying ratios of DE (phantom-like) and ordinary matter. The energy conditions and gravitational wave echoes for DESs were analyzed in \cite{sakti2021dark}, where the phantom-field contribution has been found to delay these wave echoes. Ghezzi \cite{ghezzi2011anisotropic} investigated the mass-radius ($M-R$) relation of DESs composed of isotropic ordinary matter and anisotropic DE. Some studies \cite{tudeshki2022dark, tudeshki2024effect, das2024effect} have examined how the rainbow function significantly influences the DES configurations. In Ref.~\cite{smerechynskyi2021impact}, the authors analyzed the influence of quintessence DE, described by a dynamical scalar field, on NSs. Panotopoulos et al.~\cite{panotopoulos2021slowly} investigated slowly rotating DESs consisting of isotropic matter and described by the CG EoS. Additional features such as radial pulsations, moment of inertia, and tidal deformability of DESs have been effectively explored by Pretel \cite{pretel2023radial}. This single-phase DE model has recently been extended to a hybrid setting \cite{Pretel2024PRD1, Pretel2024PRD2, 2024arXiv241213568P}, where DE is confined into the core of the NS while the crust contains ordinary matter. It was reported that an increase in the energy density jump leads to an increase in the radial stability of the NS with a DE core. The reader can also find numerous research articles related to DESs in \cite{chan2009star, bhar2018anisotropic, estevez2021chaplygin, das2023dark, bhar2023dark, das2024acceptable, das2025study, Onofrio2025}.

In addition to examining the $M-R$ relations, we focus on several key physical properties, including compactness $C$, moment of inertia $I$, tidal deformability $\Lambda$, gravitational binding energy $E_g$, and $f$-mode nonradial pulsation frequency. We will also explore several universal relations (URs) that connect these global properties, i.e., connections collectively written as $C-I-\Lambda-E_g-f$ URs. Yagi and Yunes first established the UR known as the $I–{\rm Love}–Q$ relation (where $Q$ is the quadrupole moment) for slowly rotating and tidally deformed NSs \cite{yagi2013love}, as well as for anisotropic NSs \cite{yagi2015love}. Numerous subsequent studies have investigated the URs among various macroscopic properties of compact objects \cite{haskell2013universality, chakrabarti2014q, gupta2017love, bandyopadhyay2018moment, jiang2020psr, yeung2021love, Zhao2022, Pretel2024PS, Negreiros2025}. These relations arise because certain physical quantities are strongly correlated and remain nearly independent of the star’s internal composition. As a result, EoS-insensitive correlations are essential for inferring unobservable properties from those that are measurable. Using different anisotropic models, the $I–{\rm Love}–f–C$ URs for anisotropic NSs have also been investigated \cite{das2022love, mohanty2024impact, Guedes2026}. Very recently, the URs involving fundamental modes in two-fluid DM admixed NSs have been examined under the relativistic Cowling approximation \cite{Sotani2025}. Motivated by these works, the primary goal of our analysis is to evaluate how these properties vary with the MCG parameter $\alpha$ and whether the URs hold in the context of DESs.

This work is organized as follows: In Sec.~\ref{section2}, we have reported discussions of several key features corresponding to our proposed DESs step-by-step like mass-radius relation, moment of inertia, tidal deformability, gravitational binding energy, and nonradial oscillations. Notably, in this section, we have proposed the DES as a hypothetical, static, spherically symmetric stellar configuration composed of isotropic DE fluid, and have solved the TOV equations in terms of the MCG parameter $\alpha$. In Sec.~\ref{section3}, we have presented our main goal; the $C-I-\Lambda-E_g-f$ URs for DESs under the influence of MCG EoS. We have also examined how the measurement of tidal deformability from the gravitational-wave event GW170817 can be employed to predict the canonical properties of a $1.4\,M_\odot$ compact star, thereby placing constraints on the proposed DES models in that section. Finally, the main conclusions drawn from the entire discussion are summarized in Sec.~\ref{section4}. Furthermore, to provide a clearer understanding of how the MCG EoS influences the various macroscopic properties and the URs of our proposed DES models, we present a concise yet comprehensive discussion in the Appendix \ref{extended}. This section revisits and elaborates on the analyses originally introduced in Secs.~\ref{section2} and \ref{section3}, specifically focusing on the effects of varying the parameters $A$ and $B$. In this study, we use geometrized units in which $G = c = 1$, where $G$ and $c$ are the gravitational constant and the speed of light, respectively.

\section{stellar structure equations}\label{section2}

\begin{figure}
\includegraphics[width=8.3cm]{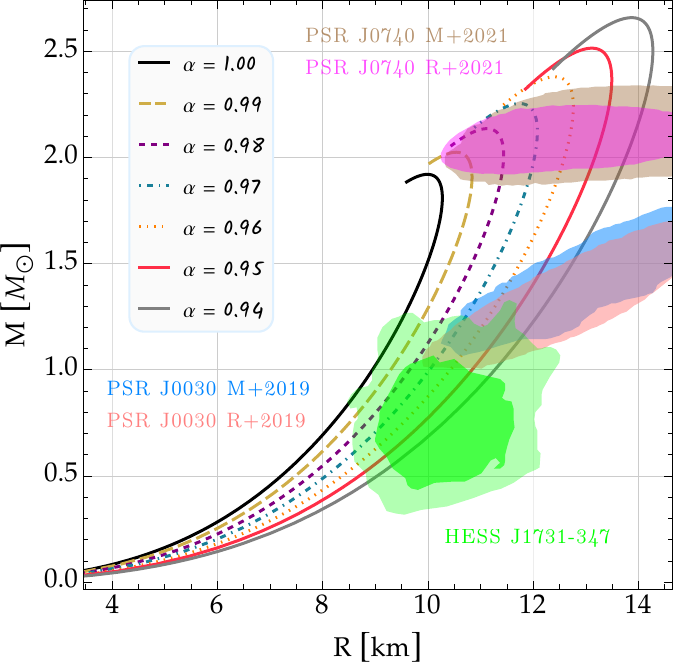}
    \caption{Mass-radius diagram for DESs with MCG EoS employing $A=0.3$, $B= 6.0\times 10^{-20}\, \rm m^{-2(1+\alpha)}$ and several values of $\alpha$, where we have included different constraints derived from compact-star observations. The green areas represent the $1\sigma$ and $2\sigma$ confidence levels for the supernova remnant HESS J1731-347 \cite{Doroshenko2022}. The pink and blue contours depict the $90\%$ CI regions obtained from the pulsar PSR J0030+0451 \cite{Riley2019, Miller2019}, while magenta and brown zones correspond to PSR J0740+6620 \cite{Riley2021, Miller2021}. One can observe that the main consequence of $\alpha$ is to increase the maximum mass as well as the radius of the stars as it decreases from $\alpha=1$. Furthermore, the DESs in this work can meet several observational measurements if $\alpha \lesssim 0.99$. }%
    \label{FigMR}%
\end{figure}

\subsection{Mass-radius relation}

To describe the spacetime geometry of our DE stellar system, we consider a static and spherically symmetric metric as usual, namely
\begin{equation}\label{LineEleEq}
ds^{2}=-e^{2\Phi(r)}dt^{2} + e^{2\Psi(r)}dr^{2} + r^{2}\left(d\theta^{2}+\sin^{2}\theta d\phi^{2}\right),
\end{equation}
where the metric potentials $\Phi(r)$ and $\Psi(r)$ depend on the radial coordinate $r$. The mass function $m(r)$, which quantifies the amount of mass within a sphere of radius $r$, is related to $\Psi(r)$ via $e^{-2\Psi(r)}= 1-2m(r)/r$. The surface radius $R$ is defined as the radial coordinate where the pressure vanishes, i.e., $p(R)=0$. The total gravitational mass $M$ of the stellar object is then given by $M=m(R)$.

In addition, to model compact stars, a relativistic perfect fluid has to be assumed. Here the energy-momentum tensor to describe an isotropic perfect fluid is
\begin{equation}\label{EMTensorEq}
T_{\mu\nu}= (\rho +p)u_{\mu}u_{\nu}+ pg_{\mu\nu}, 
\end{equation}
with $\rho$, $p$ and $u_\mu$ being the energy density, pressure
and the four-velocity of the fluid, respectively. 

In view of Eqs.~\eqref{LineEleEq} and \eqref{EMTensorEq}, the Einstein field equations therefore provide the well-known Tolman-Oppenheimer-Volkoff (TOV) equations \cite{Oppenheimer:1939ne, Tolman:1939jz}:
\begin{align}
    \frac{dm}{dr} &= 4\pi r^{2}\rho,  \label{TOV1}  \\
    \frac{dp}{dr} &= (\rho +p)\frac{m+4\pi r^{3}p}{r(2m-r)}.  \label{TOV2}
\end{align}
So far, with these two stellar structure equations it is possible to determine the mass-radius $(M-R)$ relations using the initial conditions $\rho(0)= \rho_c$ and $m(0)= 0$. However, other astrophysical observables such as moment of inertia and tidal deformability require determining the metric function $\Phi(r)$. The differential equation governing $\Phi(r)$ arises from the conservation law $\nabla_\mu T_1^\mu =0$, i.e.,
\begin{equation}\label{TOV3}
\frac{d\Phi}{dr}= -\frac{1}{\rho +p}\frac{dp}{dr}, 
\end{equation}
which must satisfy $e^{2\Phi}=e^{-2\Psi}=1-2M/R$ due to the Schwarzschild exterior solution.

Figure \ref{FigMR} shows our $M-R$ results for various values of $\alpha$, obtained by solving the TOV equations \eqref{TOV1} and \eqref{TOV2} with $\rho_c$ being an input. One observes that the main consequence of $\alpha$ (as it decreases) is to shift the $M-R$ curves to the upper right indicating that the maximum-mass values undergo a substantial increase with respect to the standard case $\alpha= 1$. This behavior is due to the fact that the DE stellar configuration supports a higher pressure as $\alpha$ decreases, as evidenced in the upper panel of Fig.~\ref{FigEoSandSpeedSound}. For $\alpha \lesssim 0.99$, the MCG EoS allows radii and masses potentially consistent with the different observational measurements such as the central compact object within the supernova remnant HESS J1731-347 \cite{Doroshenko2022}, and the millisecond pulsars PSR J0030+0451 \cite{Riley2019, Miller2019} and PSR J0740+6620 \cite{Riley2021, Miller2021}. As a consequence, while a perfect EoS is still elusive, the MCG provides significant implications for the astrophysics of compact stars made of DE.

\begin{figure*}
\includegraphics[width=8.05cm]{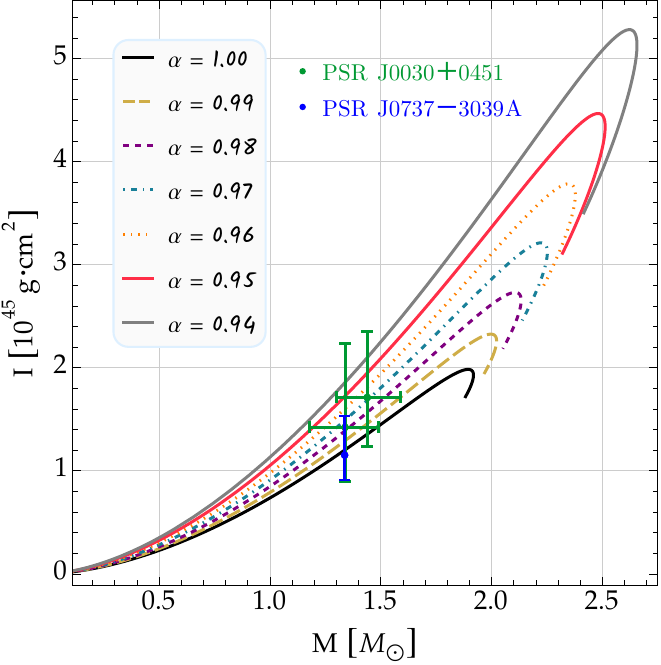}
\hspace{2mm}
\includegraphics[width=8.5cm]{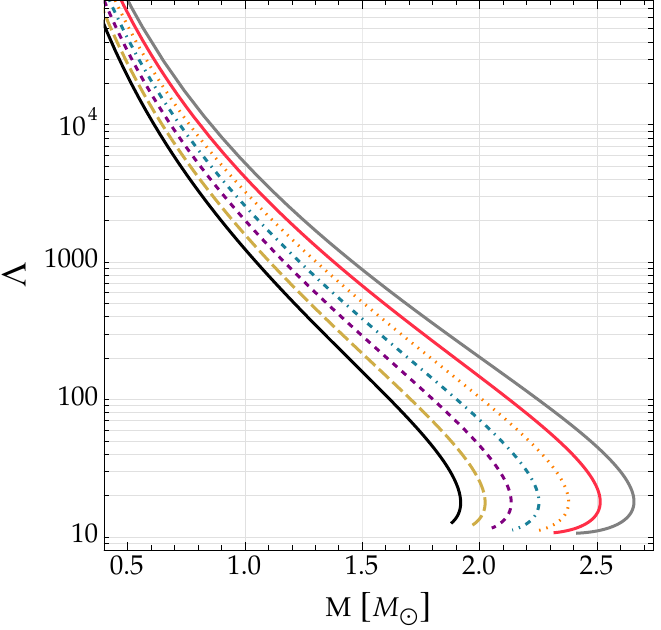}
    \caption{Left panel: Moment of inertia versus gravitational mass for the DESs presented in Fig.~\ref{FigMR}. Green dots with their respective error bars stand for the inferred properties for PSR J0030+0451 \cite{Silva2021}, while the blue bar represents the  moment of inertia of PSR J0737-3039A \cite{Landry2018}. Right panel: Tidal deformability as a function of mass. }%
    \label{FigIMLM}%
\end{figure*}

\subsection{Moment of inertia}

The moment of inertia (MoI) is a measure of the resistance of an object to changes in its rotational motion. In other words, a larger MoI means that it is harder to speed up or slow down the rotation of an object, requiring greater torque. In Newtonian physics, the MoI of a solid sphere about an axis through its center is $I= (2/5)MR^2$, where $M$ is the mass of the sphere and $R$ is its radius. For example, for a compact star of $2\, M_\odot$ and a radius of $10\, \rm km$, the MoI is $I\sim 10^{45}\, \rm g\cdot cm^2$. Although this approximation for $I$ is somewhat crude, it already gives us an idea of its order of magnitude for spherically symmetric stars, as we will see later. Of course, since we are dealing with the relativistic structure of compact stars, it becomes necessary to use a relativistic version for the MoI. Although direct measurements of the MoI are challenging, researchers use various techniques like pulsar timing and gravitational wave observations to constrain the values of $I$ \cite{Landry2018, Hu2024}. Such measurements, combined with mass and radius data, can help refine our understanding of dense matter at extreme densities and test modified gravity theories. Therefore, it is of utmost importance to calculate the MoI for our DE stellar models, and, as we shall see later, this astrophysical observable plays a crucial role when examining URs.

In this subsection, we compute the MoI of a DES in the context of the slowly rotating approximation \cite{hartle1967slowly}. Specifically, by considering rotational corrections only up to first order in the angular velocity of the star $\Omega$, the spacetime metric (\ref{LineEleEq}) is replaced by its corresponding slowly rotating form, namely:
\begin{align}\label{MIlineele}
ds^{2}=& -e^{2\Phi(r)}dt^{2} + e^{2\Psi(r)}dr^{2} + r^{2}\left(d\theta^{2}+\sin^{2}\theta d\phi^{2}\right)\nonumber \\
&- 2\omega(r,\theta)r^{2}\sin^{2}\theta dt d\phi.  
\end{align}
Remark that both $\Phi(r)$ and $\Psi(r)$ remain functions solely of the radial coordinate $r$, while $\omega(r, \theta)$ denotes the angular velocity of the local inertial frames induced by the rotation of the DES. It should be noted that $\Omega$ represents the angular velocity of the stellar fluid as measured by an observer at rest at a given spacetime point $(t, r, \theta, \phi)$. Thus, the four-velocity vector of the fluid up to linear order in $\Omega$, can be expressed as $u^{\mu} = \left(e^{-\Phi},\ 0,\ 0,\ \Omega e^{-\Phi} \right)$.

As introduced in Ref.~\cite{hartle1967slowly}, here it is useful to define the quantity $\varpi = \Omega- \omega$, and the $03-$component of the field equations for the metric (\ref{MIlineele}) contributes an additional differential equation for the angular velocity. By retaining only first-order terms in $\Omega$, this component reads
\begin{eqnarray}
\frac{e^{\Phi-\Psi}}{r^{4}}\frac{\partial}{\partial r}\left[e^{-\Phi-\Psi} r^{4}\frac{\partial \varpi}{\partial r}\right]+ \frac{1}{r^{2}\sin^{3}\theta}\frac{\partial}{\partial\theta}\left[\sin^{3}\theta\frac{\partial \varpi}{\partial r}\right]\nonumber\\
=16\pi (\rho +p)\varpi .
\end{eqnarray}
Indeed, one can demonstrate that $\varpi$ depends solely on the radial coordinate $r$ and can be determined by solving the following differential equation (refer to Ref.~\cite{pretel2023radial} for more details)
\begin{equation}\label{MI9}
\frac{e^{\Phi-\Psi}}{r^{4}}\frac{d}{dr}\left[e^{-\Phi-\Psi} r^{4}\frac{d\varpi}{dr}\right]= 16\pi(\rho +p)\varpi .
\end{equation}

Note that the metric functions have already been calculated by means of the TOV equations, which still hold to rotational corrections at first order in the angular velocity. The relativistic MoI for a slowly rotating DES can then be explicitly expressed as:
\begin{equation}\label{MI10}
I= \frac{8\pi}{3}\int_{0}^R (\rho+p)e^{\Psi-\Phi}r^{4}\left(\frac{\varpi}{\Omega}\right)dr ,
\end{equation}
where the boundary conditions for the above differential equation \eqref{MI9} are naturally imposed upon specifying an arbitrary central value $\varpi(0)$:
\begin{equation}
\frac{d\varpi}{dr}\bigg|_{r=0}=0, \qquad \quad \lim_{r\rightarrow \infty}\varpi=\Omega.
\end{equation}
The first condition guarantees regularity at the stellar center, while the second enforces asymptotic flatness at spatial infinity where $\omega \rightarrow 0$. After obtaining the solution $\varpi(r)$, the MoI can be calculated using the integral (\ref{MI10}). The left panel of Fig.~\ref{FigIMLM} displays the MoI for the stellar configurations shown in Fig.~\ref{FigMR}, from which we can observe that decreasing $\alpha$ results in higher MoIs, with maximum values up to $\sim 5.3\times 10^{45}\, \rm g\cdot cm^2$ for $\alpha= 0.94$. It can be seen that our theoretical calculations are highly compatible with the inferred MoI for PSR J0030+0451 \cite{Silva2021} and PSR J0737-3039A \cite{Landry2018}.

\subsection{Tidal deformability}

In this subsection, we analyze the dimensionless tidal deformability for our DES models. Such tidal deformability characterizes how a compact star responds to an external tidal field $\mathcal{E}_{ij}$ created by its distant companion star. Consequently, the originally spherically symmetric star acquires an induced quadrupole moment $\mathcal{Q}_{ij}$ and, to linear order in $\mathcal{E}_{ij}$, this deformation can be expressed as follows
\begin{equation}
\mathcal{Q}_{ij}= \lambda\mathcal{E}_{ij},
\end{equation}
where $\lambda$ is tidal deformability of the compact star and is related to the tidal Love number $k_{2}$ via
\begin{equation}
\lambda= \frac{2}{3}k_{2}R^{5}.
\end{equation}

To define the dimensionless tidal deformability, we introduce the quantity $\Lambda= \lambda/M^5= 2k_2C^{-5}/3$, where $C$ is the compactness of the DES. Accordingly, our subsequent analysis will focus on $\Lambda$. To evaluate the Love number $k_{2}$, we follow the approach of Thorne and Campolattaro \cite{thorne1967non} by examining linear perturbations of the background metric:
\begin{equation}
g_{\mu\nu}=g^{0}_{\mu\nu}+h_{\mu\nu},
\end{equation}
with $g^{0}_{\mu\nu}$ being the equilibrium configuration metric and $h_{\mu\nu}$ is a linearized metric perturbation. So, the perturbed metric can be written as \cite{hinderer2008tidal}
\begin{eqnarray}
h_{\mu\nu}= \text{diag}\left[-e^{2\Phi(r)}H_{0}, e^{2\Psi(r)}H_{2}, r^{2}K, r^{2}K\sin^{2}\theta\right]\nonumber\\
\times Y_{2m}(\theta,\phi),
\end{eqnarray}
where the variables $H_0$, $H_2$, and $K$ are radial functions determined by the perturbed Einstein equations. The perturbed energy-momentum tensor is given by 
$\delta T^{\nu}_{\mu}=\text{diag}\left(-\delta\rho, \delta p, \delta p, \delta p\right)$. These fluid and metric perturbations are substituted into the linearized Einstein equations to read
\[
\begin{cases}
 H_{0}=-H_{2}= H & \text{from }\ \delta G_{2}^{2}-\delta G_{3}^{3}=0, \\
 K'=2H\Phi'+H' & \text{from }\  \delta G_{1}^{2}=0, \\
 \delta p=\frac{H}{8\pi r}e^{-2\Psi}(\Psi' +\Phi')Y_{2m} & \text{from }\  \delta G_{2}^{2}=8\pi \delta p .
\end{cases}
\]

In addition, from $\delta G_0^0 - \delta G_1^1 =-8\pi(\delta\rho+\delta p)$, we obtain a differential equation for the perturbation variable $H(r)$
\begin{align}
&H''+ H'\left\lbrace\frac{2}{r}+e^{2\Psi}\left[\frac{2m}{r^{2}}+4\pi r (p-\rho)\right]\right\rbrace + \nonumber  \\
&H\left\lbrace e^{2\Psi}\bigg[ \frac{4\pi(\rho+p)}{dp/d\rho} + 4\pi(5\rho + 9p)- \frac{6}{r^2} \bigg] -4 \Phi'^{2}\right\rbrace 
=0.
\end{align}

\begin{figure*}
\includegraphics[width=8.33cm]{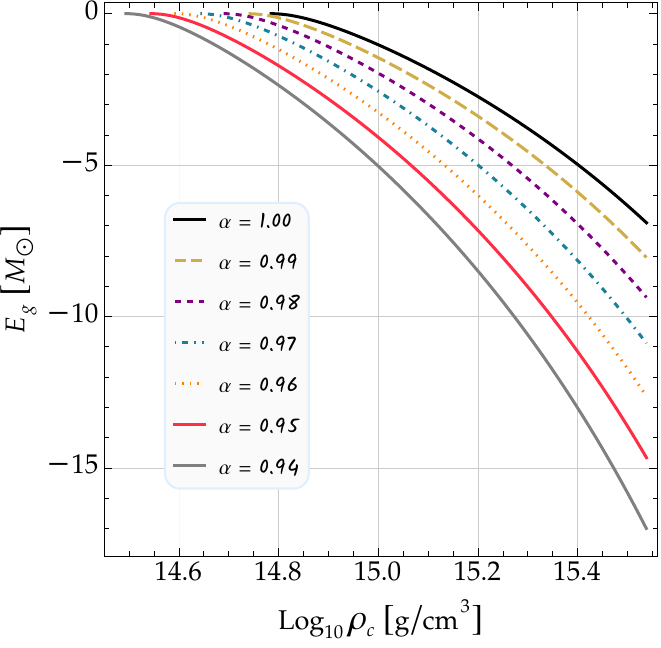}
\hspace{2mm}
\includegraphics[width=8.25cm]{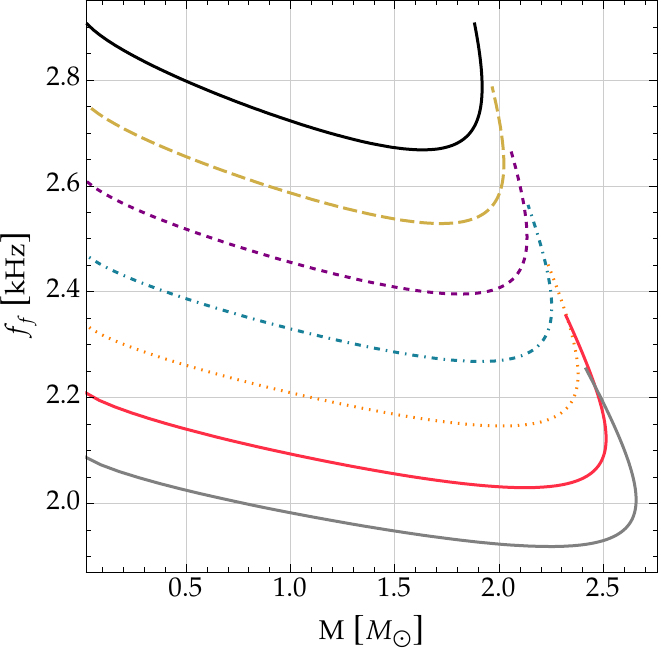}
    \caption{Left: Relation between the gravitational binding energy and central density. Right: Fundamental nonradial oscillation frequency with respect to gravitational mass. Given a mass $M$, the decrease in $\alpha$ leads to a decrease in the $f$-mode frequency. }%
    \label{FigEgRhocffMass}%
\end{figure*}

The tidal Love number is obtained by matching the internal and external solutions of the perturbed variable $H$ at the stellar surface $r = R$, namely \cite{hinderer2008tidal, Hinderer2010}
\begin{align}\label{LoveNumEq}
    k_2 =&\ \frac{8}{5}(1- 2C)^2C^5 \left[ 2C(\beta -1) - \beta+ 2 \right]  \nonumber  \\
    &\times \left\lbrace 2C[ 4(\beta+ 1)C^4 + (6\beta- 4)C^3 \right.  \nonumber  \\
    &\left.+\ (26- 22\beta)C^2 + 3(5\beta -8)C - 3\beta+ 6 \right]   \nonumber  \\
    &\left.+\ 3(1-2C)^2\left[ 2C(\beta- 1)- \beta +2 \right]\ln(1-2C) \right\rbrace^{-1} ,
\end{align}
where $\beta= y(R)- 4\pi R^3 \rho_s/M$ with $y= rH'/H$. Note that $\rho_s$ represents the energy density at the surface of the DES. This correction term ``$-4\pi R^3 \rho_s/M$'' is included due to the fact that the energy density at the surface is finite and non-null \cite{Pretel2024}, as can be seen from the EoS \eqref{EqoS} when the pressure vanishes. In fact, for the MCG model, the surface energy density can be written as
\begin{equation}
    \rho_s = \left( \frac{B}{A} \right)^{\frac{1}{1+\alpha}} .
\end{equation}

In the right panel of Fig.~\ref{FigIMLM} we display the dimensionless tidal deformability $\Lambda$ as a function of the gravitational mass $M$. Given a fixed mass, it can be seen that decreasing $\alpha$ leads to larger $\Lambda$. In other words, the smaller the value of $\alpha$, the greater the deformation of a DES due to the tidal effect created by the companion star. On the other hand, given a fixed $\alpha$, less massive DESs have greater deformability, similar to the case of QSs and NSs.

\subsection{Gravitational binding energy}

The gravitational binding energy (GBE) has also been shown to be a characteristic quantity that can reveal information about the internal structure of a NS \cite{Jiang2019}, in addition to being correlated with other macroscopic properties through certain URs. In this regard, it is also important to calculate the GBE for our DE stellar configurations. To do so, we start with the definition of the proper mass of a compact star, given by \cite{Bagchi2011}
\begin{align}
    M_{pr} &= 4\pi\int_0^R r^2\rho(r)e^{\Psi(r)}dr  \nonumber  \\
    &= 4\pi\int_0^R r^2\rho(r) \left[ 1-\frac{2m(r)}{r} \right]^{-1/2}dr ,
\end{align}
which is a consequence of integrating the energy-density distribution $\rho(r)$ over the proper volume element in the metric \eqref{LineEleEq}. Thus, from the total gravitational mass $M$ and proper mass we can obtain the GBE as
\begin{equation}\label{GBEEq}
    E_g = M- M_{pr} .
\end{equation}

The left plot of Fig.~\ref{FigEgRhocffMass} shows the behavior of the GBE as a function of the central density. For a given $\alpha$ value, $E_g$ decreases with increasing central density. In other words, more massive DESs exhibit a greater amount of $\vert E_g \vert$. Furthermore, for a given $\rho_c$, the GBE decreases as $\alpha$ becomes smaller. We will return to this binding energy later, when we discuss URs.

\subsection{Nonradial oscillations}\label{subNonOscill}

In order to calculate the nonradial oscillation frequencies in DESs, here we also solve the first-order ordinary differential equations that govern the nonradial perturbations of relativistic compact stars in the Cowling approximation\footnote{This approximation is defined when the stellar fluid oscillates on a fixed background metric, i.e., the equations are derived assuming $\delta g^{\mu\nu}= 0$.}, namely \cite{Sotani2011}
\begin{align}
    \frac{dW}{dr} &= \frac{1}{v_s^2}\left[ \frac{\nu^2r^2 V}{e^{2\Phi- \Psi}} + \Phi'W\right] - \ell(\ell+1)e^\Psi V,  \label{NonRadEq1}  \\
    \frac{dV}{dr} &= 2\Phi'V - \frac{e^\Psi W}{r^2},  \label{NonRadEq2}
\end{align}
where $W(r)$ and $V(r)$ are perturbative variables, initially written as $W(t,r)= W(r)e^{i\nu t}$ and $V(t,r)= V(r)e^{i\nu t}$ in the Lagrangian displacement vector
\begin{equation}
\xi^{i}= \left(e^{-\Psi}W,-V\partial_{\theta},-\frac{V}{\sin^{2}\theta}\partial_{\phi}\right)\frac{Y_{\ell m}}{r^2} ,
\end{equation}
with $\nu$ being the nonradial oscillation frequency to be determined, and $Y_{\ell m}=Y_{\ell m}(\theta,\phi)$ are the spherical harmonics. As usual, in our numerical calculations we will deal with the quadripolar modes, i.e., when $\ell =2$.

The system of equations \eqref{NonRadEq1} and \eqref{NonRadEq2} is nothing more than an eigenvalue problem where the aim is to calculate mainly the eigenfrequencies of the fundamental mode $\nu_f$, corresponding to the lowest order nonradial oscillation mode, which is characterized by the absence of nodes. It then becomes necessary to impose appropriate boundary conditions at the stellar center and at the stellar surface. At $r=0$, the radial variables $W$ and $V$ assume the form
\begin{align}\label{CenCond_NRO}
W &= cr^{\ell+1},  &  V &= -c\frac{r^{\ell}}{\ell},
\end{align}
where $c$ depicts a dimensionless constant. Additionally, at $r= R$, the following condition must be obeyed
\begin{equation}\label{SurCond_NRO}
    \frac{\nu^2 V}{e^{2\Phi}} + \frac{\Phi'W}{e^{\Psi}r^2} =0 ,
\end{equation}
which is the result of imposing that the Lagrangian perturbation of pressure be zero at the surface of the star.

In the right plot of Fig.~\ref{FigEgRhocffMass} we show our results for the fundamental nonradial oscillation frequency $f_f= \nu_f/2\pi$. One observes that, given a fixed $\alpha$, the $f$-mode frequency decreases to a certain minimum value and then grows with increasing gravitational mass. Furthermore, the frequency $f_f$ decreases drastically as the parameter $\alpha$ becomes smaller. We must emphasize that, for the $f$-mode frequencies, the typical error level brought by the Cowling approximation is around $20\%$ compared to the full GR calculations \cite{Zhao2022, Sotani2020PRD, Kunjipurayil2022}.

\begin{figure*}
\includegraphics[width=8.48cm]{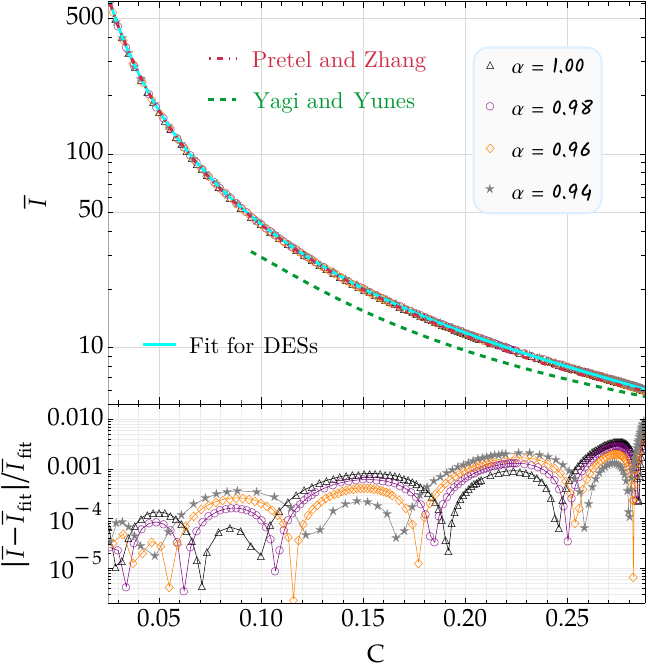}
\hspace{2mm}
\includegraphics[width=8.33cm]{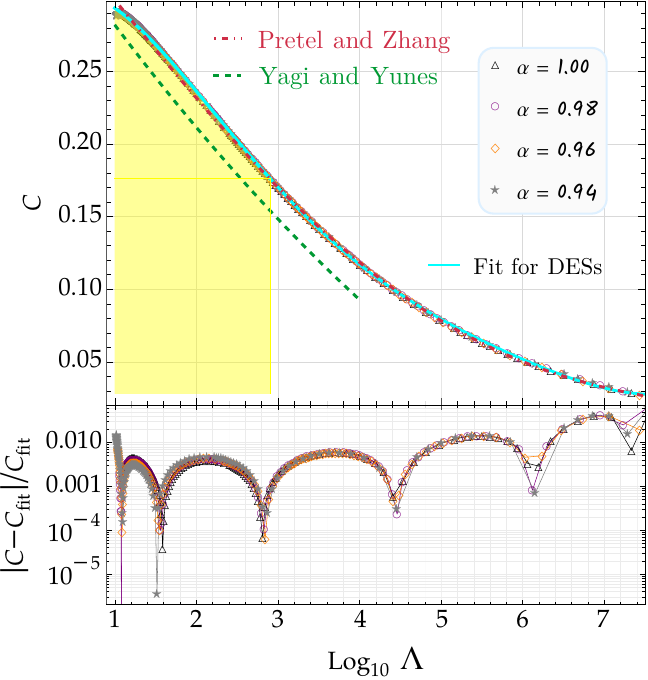}
    \caption{Left: Normalized moment of inertia $\bar{I}$ as a function of the compactness $C$, where our polynomial function \eqref{UREqIC} is represented by the cyan curve. Right: Compactness versus tidal deformability relation, where our fit for DESs \eqref{UREqCLambda} is shown in cyan. For both URs we have considered four values for the CDF parameter $\alpha$, and the fit residuals are displayed in the lower plots. The red dot-dashed lines are the fitting functions for isotropic interacting QSs derived in \cite{Pretel2024}, while the green dashed curves are the analytical expressions for NSs \cite{YagiYunes2017}. Additionally, the filled yellow area stands for the EoS-independent constraint $\Lambda_{1.4} \leq 800$ from the GW170817 signal \cite{Abbott2017PRL}. }
    \label{FigICCL}%
\end{figure*}

\section{Universal relations}\label{section3}

According to Refs.~\cite{yagi2013love, yagi2015love, YagiYunes2017, chan2016universality, breu2016maximum}, the macroscopic properties of a compact star, such as its compactness, MoI and tidal deformability are related through URs. In that regard, here we examine various types of URs involving these astrophysical observables, and we will also consider the GBE and the $f$-mode nonradial oscillation frequency; all of which have been previously defined, i.e., $C-I-\Lambda-E_g-f$ URs. Our primary goal is to investigate these URs for DESs under the influence of the MCG EoS parameter $\alpha$. We will compare our UR results with those already known for NSs and QSs.

We begin by examining the $I-C$ correlation, originally proposed in \cite{ravenhall1994neutron} and later thoroughly investigated by Breu and Rezzolla \cite{breu2016maximum} within the framework of GR. Subsequent studies have extended and modified this relation across various contexts, including the double pulsar system using higher-order polynomial fitting \cite{lattimer2005constraining}, strange stars \cite{bejger2002moments}, NSs in scalar-tensor theories and $R$-squared gravity \cite{staykov2016moment,popchev2019moment}, and QSs within the astrophysical context of $f(R,T,L_m)$ theories \cite{Pretel2024PS}. Here, therefore, we explore the $I-C$ relation specifically for isotropic DESs. Breu and Rezzolla \cite{breu2016maximum} highlighted the universal nature between the normalized MoI $\bar{I} = I/ M^{3}$ and the compactness $C=M/R$. Such relation is expressed via the following polynomial form, as discussed in \cite{Pretel2024},
\begin{equation}\label{UREqIC}
  \log_{10}\bar{I} = \sum_{n=-3}^3 a_n C^n ,
\end{equation}
where $a_{n}$ are the fitting coefficients and are reported in $1^{\text{st}}$ column of Table \ref{table1}. The left panel of Fig.~\ref{FigICCL} displays our empirical function (\ref{UREqIC}) graphically by a cyan solid line. Interestingly, the $\bar{I}-C$ relation for DESs shows a marked deviation from that of NS matter \cite{YagiYunes2017}, illustrated by the green dashed curve, while remaining consistent with the behavior found for interacting quark stars (IQSs) \cite{Pretel2024}, shown by the red dot-dashed curve. Furthermore, in the bottom panel of that figure, we present the fit residuals, defined as $|\bar{I} - \bar{I}_{\text{fit}}|/\bar{I}_{\text{fit}}$, which are nothing more than relative fractional errors between the fitting function and the numerical results for each value of the CDF parameter $\alpha$. With a maximum fractional difference for the DES sequences of around $1\%$ in the range of adopted $\alpha$, our findings reveal that the $\bar{I}-C$ correlation retains its approximate universality. This implies that the relation remains largely unaffected by variations in $\alpha$.

The right panel of Fig.~\ref{FigICCL} illustrates the $C-\Lambda$ UR corresponding to the fit function as follows:
\begin{equation}\label{UREqCLambda}
    C = \sum_{n=0}^5 b_n \left( \log_{10}\Lambda \right)^n ,
\end{equation}
with the fitting parameters $b_{n}$ given in Table \ref{table1}. The  relative fractional errors are computed using the expression $|C - C_{\text{fit}}|/C_{\text{fit}}$ and are shown in the lower right panel of Fig.~\ref{FigICCL}. For comparison reasons, the fits corresponding to NSs \cite{YagiYunes2017} and IQSs \cite{Pretel2024} were included by green dashed and red dot-dashed lines, respectively. Our findings show that the $C-\Lambda$ relation is consistent with that of QSs, but differs substantially from the NS case. In other words, we can distinguish DESs from NSs, but not from QSs, in the $C-\Lambda$ UR. As displayed in the bottom panel, the fractional deviation between the data and the empirical formula (\ref{UREqCLambda}) indicates that such relation exhibits approximate universality, with maximum deviations that reach the $\sim5\%$.

\begin{figure}
    \centering
    \includegraphics[width=8.3cm]{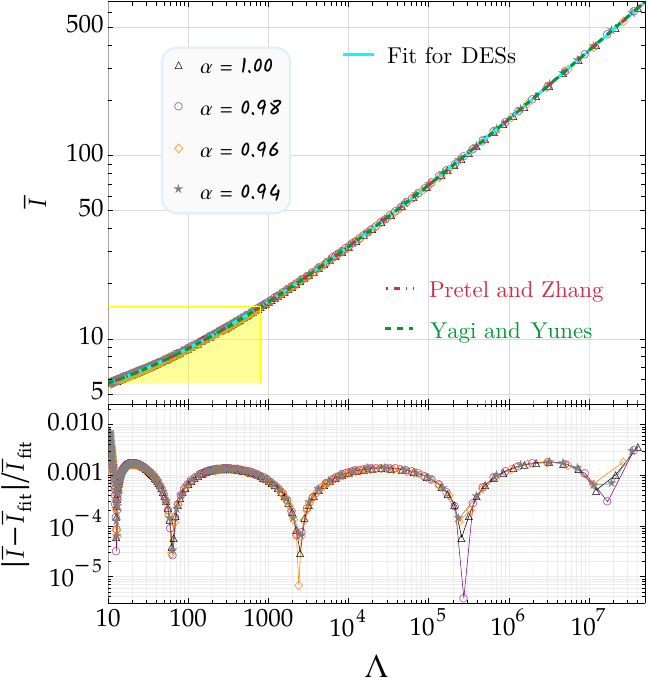}
    \caption{Normalized MoI versus tidal deformability relation for our DE configurations, where the cyan curve indicates the polynomial expression \eqref{UREqILambda}. The yellow region and the red and green curves represent the same as in Fig.~\ref{FigICCL}. }
    \label{FigIL}
\end{figure}

Continuing with our UR analysis, we now examine the relationship between MoI and tidal deformability, commonly known in the literature as the $I-{\rm Love}$ correlation or simply $I-\Lambda$ relation. Specifically, we compute this correlation for isotropic DESs using the normalized MoI, defined as $\bar{I} = I / M^{3}$~\cite{breu2016maximum}. For a given choice of the model parameter $\alpha$, we plot $\bar{I}$ as a function of the dimensionless tidal deformability $\Lambda$ (see Fig.~\ref{FigIL}). To quantitatively describe this relation, we perform a fitting using the power-series expansion
\begin{equation}\label{UREqILambda}
\log_{10}\bar{I} = \sum_{n=0}^{4}c_{n}\left(\log_{10}\Lambda\right)^{n},
\end{equation}
where the coefficients $c_{n}$ are listed in the $3^{\text{rd}}$ column of Table \ref{table1}. The empirical formula \eqref{UREqILambda} for our DE configurations is represented by the cyan curve in the upper panel of Fig.~\ref{FigIL}. Besides, the lower panel illustrates the absolute fractional deviation of the numerical data with respect to the fitted values, given by the expression $|\bar{I} - \bar{I}_{\text{fit}}|/\bar{I}_{\text{fit}}$. One observes that the considered variations in $\alpha$ lead to fractional errors smaller than $1\%$, indicating a strong universal correlation between the two astrophysical observables. Just like in the previous URs, the fitting curve proposed by Pretel and Zhang \cite{Pretel2024} for IQSs is shown using a red dot-dashed line. For comparison, the fitting function for NSs found by Yagi and Yunes~\cite{YagiYunes2017} is also displayed, represented by a green dashed line. We notice that the $\bar{I}-\Lambda$ correlation for IQSs and NSs is nearly indistinguishable from that of DESs with MCG EoS.

\begin{table*}
\caption{\label{table1} Fitting coefficients for the empirical formulas \eqref{UREqIC}, \eqref{UREqCLambda}, \eqref{UREqILambda}, \eqref{UREqfC}, \eqref{UREqfI} and \eqref{UREqfLambda}.  }
\begin{ruledtabular}
\begin{tabular}{c|c|c|c|c|c}
$I-C$ UR  &  $C-\Lambda$ UR  &  $I-\Lambda$ UR  &  $f-C$ UR  &  $f-I$ UR  &  $f-\Lambda$ UR  \\
\hline
  $a_{-3} [10^{-5}]= -7.3471$  &  $b_0 [10^{-1}]= 2.8901$  &  $c_0 [10^{-1}]= 6.3943$  &  $d_0 [10^{-3}]= -2.3720$  &  $g_0 [10^{-2}]= -6.6982$  &  $h_0 [10^{-1}]= 1.2344$  \\
  $a_{-2} [10^{-1}]= 4.0739$  &  $b_1 [10^{-2}]= 6.2654$  &  $c_1 [10^{-2}]= 7.3393$  &  $d_1 [10^{-1}]= 2.4688$  &  $g_1= 2.2555$  &  $h_1 [10^{-1}]= 1.3959$  \\
  $a_{-1} [10^{-2}]= 2.6048$  &  $b_2 [10^{-2}]= -7.5778$  &  $c_2 [10^{-2}]= 4.6025$  &  $d_2= 2.3994$  &  $g_2 [10^{1}]= -2.5727$  &  $h_2 [10^{-1}]= -1.2917$  \\
  $a_0 = 5.2234$  &  $b_3 [10^{-2}]= 1.9697$  &  $c_3 [10^{-3}]= -2.8268$  &  $d_3= -4.5439$  &  $g_3 [10^{2}]= 1.7079$  &  $h_3 [10^{-2}]= 4.3981$  \\
  $a_1 [10^{1}]= -4.2518$  &  $b_4 [10^{-3}]= -2.2143$  &  $c_4 [10^{-5}]= 5.4500$  &  $d_4= 1.3013$  &  $g_4 [10^{2}]= -5.8828$  &  $h_4 [10^{-3}]= -7.8692$  \\
  $a_2 [10^{2}]= 1.7695$  &  $b_5 [10^{-5}]= 9.3889$  &  $--$  &  $--$  &  $g_5 [10^{3}]= 1.0290$  &  $h_5 [10^{-4}]= 7.3554$  \\
  $a_3 [10^{2}]= -2.7516$  &  $--$  &  $--$  &  $--$  &  $g_6 [10^{2}]= -7.2282$  &  $h_6 [10^{-5}]= -2.8200$  \\
\end{tabular}
\end{ruledtabular}
\end{table*}

\begin{figure*}
\includegraphics[width=8.4cm]{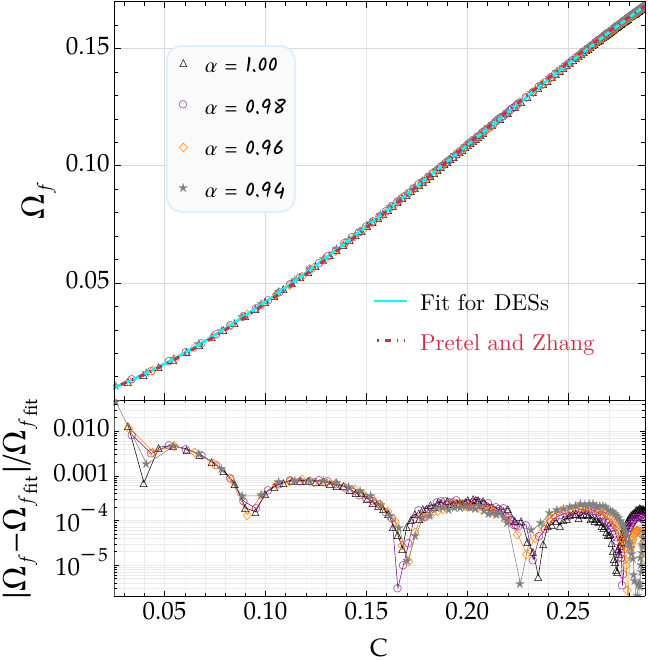}
\hspace{2mm}
\includegraphics[width=8.354cm]{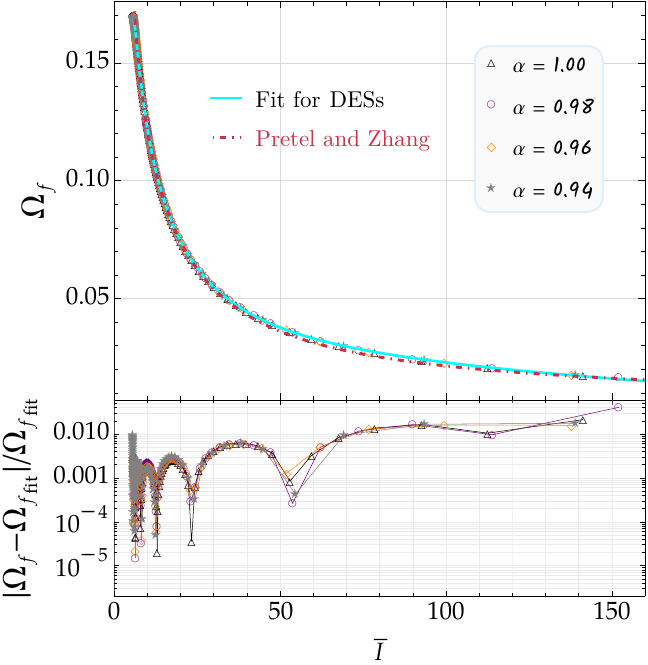}
    \caption{Dimensionless $f$-mode frequency $\Omega_f$ as a function of the compactness (left) and normalized MoI (right) for DESs. The cyan curves representing our fit functions are similar to those obtained for isotropic IQSs \cite{Pretel2024}. As in the other URs, the relative fractional difference between the numerical results and the fitting curve are presented in the lower panels. According to these relations, it is not possible to distinguish IQSs from DESs. }
    \label{FigfCfI}%
\end{figure*}

We now investigate the relation between the fundamental nonradial oscillation frequency and the compactness of DESs made of CDF. For this purpose, here it is convenient to define the dimensionless $f$-mode frequency as $\Omega_f = M \nu_f$, where $\nu_f$ has already been calculated above. In the left panel of Fig.~\ref{FigfCfI}, we present such $f-C$ relation, in which the cyan curve corresponds to the following fit function
\begin{equation}\label{UREqfC}
\Omega_f = \sum_{n=0}^{4}d_{n}C^n,
\end{equation}
where $d_{n}$ are the fitting coefficients, reported in the $4^{\text{th}}$ column of Table \ref{table1}. Remarkably, one of the key applications of the $f-C$ UR lies in inferring the mass and radius of compact stars from the observed oscillation mode data, as demonstrated by Andersson and Kokkotas \cite{Andersson:1997rn}. According to this reference, the $f-C$ UR can be utilized to predict $M$ and $R$ with surprising precision. Now, it is evident that the $f-C$ relation is indistinguishable from the curve obtained for interacting quark matter \cite{Pretel2024} and furthermore such a relation is insensitive to variations of $\alpha$ in compact stars made of DE to $\sim 4\%$ level for low compactness and less than $0.1\%$ for $C\gtrsim 0.08$, as displayed in the lower left panel of Fig.~\ref{FigfCfI}.

As shown in Ref.~\cite{Lau_2010}, the $f$-mode pulsation frequency is also related to the MoI of a compact star. With this in mind, the right plot of Fig.~\ref{FigfCfI} presents the correlation between the dimensionless $f$-mode frequency $\Omega_f$ and the normalized MoI $\bar{I}$, with our numerical data fitted by the following function
\begin{equation}\label{UREqfI}
\Omega_f = \sum_{n=0}^{6}g_n \left( \frac{1}{\bar{I}} \right)^n,
\end{equation}
with the coefficients $g_{n}$ given in the $5^{\text{th}}$ column of Table \ref{table1}. It is observed that our fitting curve (\ref{UREqfI}) for DESs is indistinguishable from the empirical function obtained for isotropic IQSs \cite{Pretel2024}. Furthermore, the $f-\bar{I}$ relation remains approximately universal with maximal deviations at $\sim 4\%$ level in the range of adopted values for the MCG EoS parameter $\alpha$.

\begin{figure}
\includegraphics[width=8.3cm]{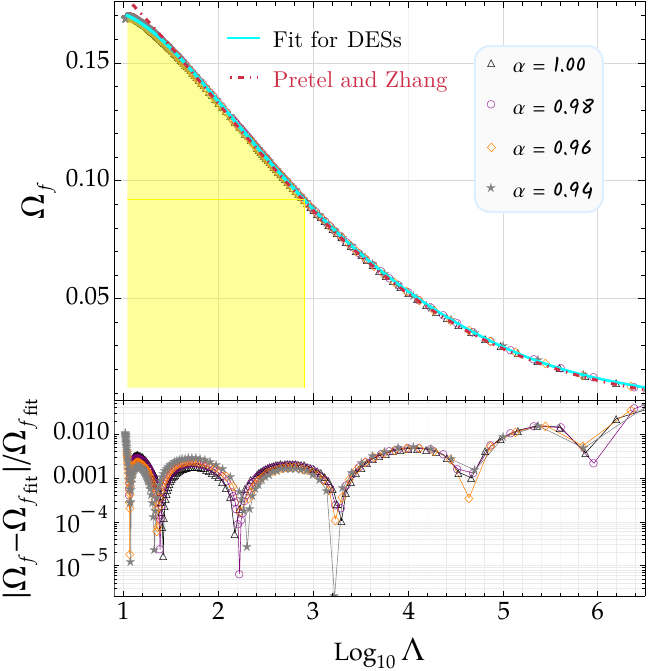}
\hspace{2mm}
    \caption{UR between the dimensionless $f$-mode frequency and tidal deformability for several values of the MCG EoS parameter $\alpha$, together with our cyan-colored power-series expansion given in Eq.~\eqref{UREqfLambda}. As in the other URs, the lower panel displays the relative fractional difference and the red dot-dashed curve corresponds to IQSs \cite{Pretel2024}. }
    \label{FigfL}%
\end{figure}

Chan and collaborators \cite{chan2014multipolar} derived EoS-insensitive empirical relations connecting the $f$-mode pulsation frequency with the tidal deformability. Consequently, we also explore this kind of universal correlations for DESs, which we refer to as the $f-\Lambda$ relation. According to Ref.~\cite{Pretel2024}, the dimensionless $f$-mode frequency $\Omega_{f}$ as a function of the tidal deformability $\Lambda$ can be written as 
\begin{equation}\label{UREqfLambda}
\Omega_f = \sum_{n=0}^{6}h_{n}\left(\log_{10}\Lambda\right)^{n},
\end{equation}
where $h_{n}$ are fitting coefficients, reported in the $6^{\text{th}}$ column of Table \ref{table1}. Similar to the above URs, one can easily observe that the $f-\Lambda$ relation is also insensitive to the variation of $\alpha$. Furthermore, the maximum fractional difference for this correlation is of the order of $\sim 4\%$. As shown in Fig.~\ref{FigfL}, this result again demonstrates that DESs cannot be distinguished from QSs using the $f-\Lambda$ UR.

\begin{figure*}
    \centering
    \includegraphics[width=5.79cm]{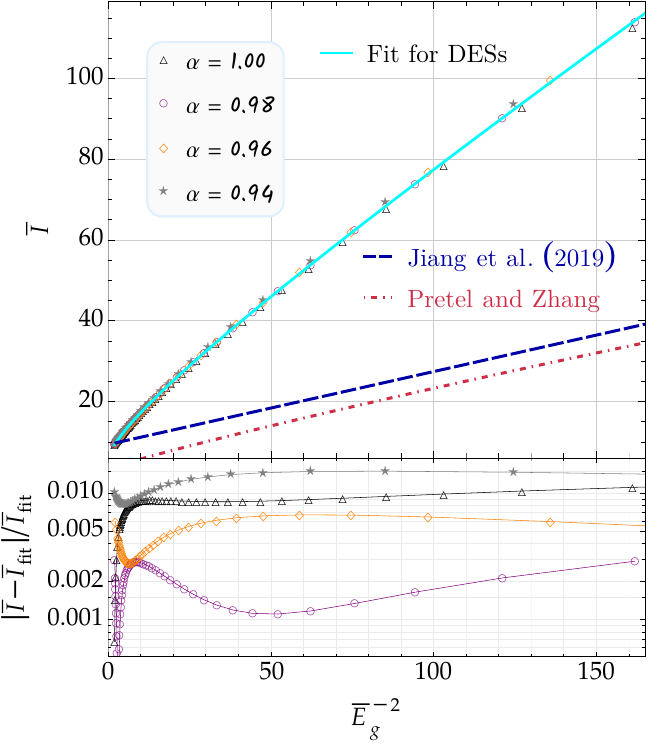}
    \includegraphics[width=6.0cm]{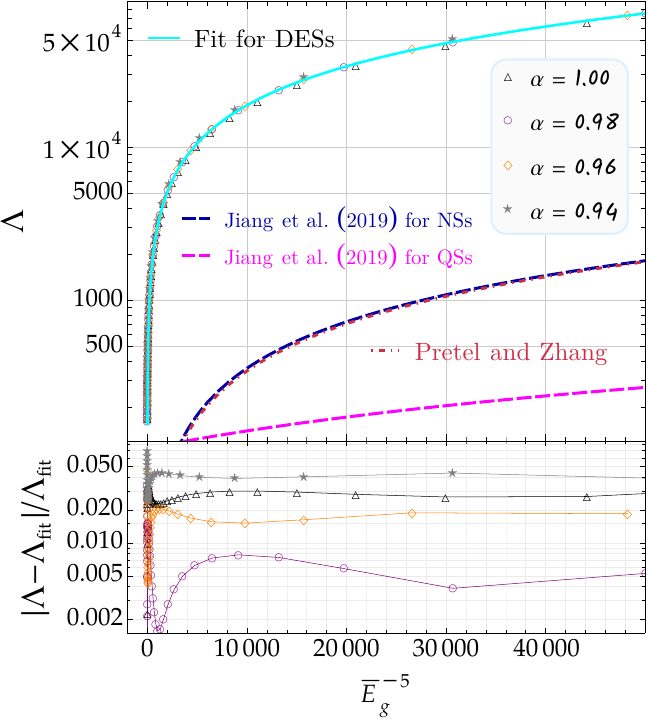}
    \includegraphics[width=5.864cm]{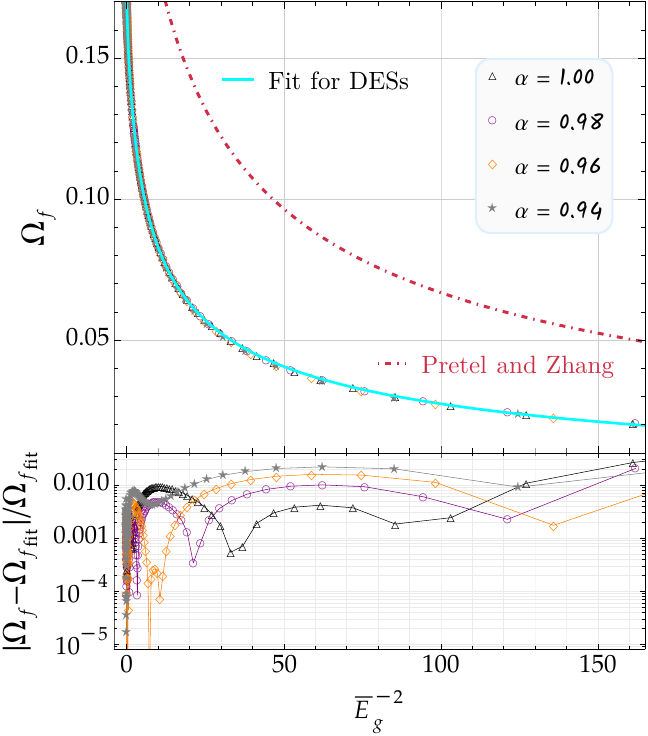}
    \caption{ Left panel: $\bar{I}-\bar{E}_g^{-2}$ UR, where the cyan line depicts the fitting function \eqref{UREqIEg} obtained for four values of the MCG EoS parameter $\alpha$. The blue dashed curve corresponds to the empirical formula for NSs and QSs reported by Jiang et at.~\cite{Jiang2019}. Middle panel: $\Lambda-\bar{E}_{g}^{-5}$ UR together with our fitted expression (\ref{UREqLEg}) shown in cyan. The blue and magenta curves denote the fitting functions for NSs and QSs, respectively, taken from Ref.~\cite{Jiang2019}. Right panel: Relation between the dimensionless $f$-mode frequency and $\bar{E}_{g}^{-2}$ for different MCG EoSs along with our empirical function \eqref{UREqfEg} displayed in cyan. In all panels, the red dot-dashed curves correspond to the fitting formulas for IQSs (see Appendix \ref{URsforIQSs}), obtained for three values of the dimensionless parameter $\tilde{\lambda}$ associated with strong-interaction effects \cite{Pretel2024}. Overall, these results demonstrate that compact objects described by DES models can be clearly distinguished from QSs (and NSs) through correlations involving the GBE.}
    \label{FigLEgfEg}
\end{figure*}

For both NSs and QSs, it has also been shown that there exists a universal behavior between the GBE and some macroscopic properties such as MoI and tidal deformability \cite{Jiang2019}. In fact, it was argued that these URs provide a potential way to estimate the GBE if the stellar mass and the MoI/tidal deformability are precisely measured. This gives us clues that there could be certain intriguing URs between the GBE and other global quantities of DESs. Therefore, in this work we focus on examining the connection between the dimensionless GBE and the normalized MoI, tidal deformability and dimensionless $f$-mode oscillation frequency of DESs. Below we give an approximate formula to describe the correlation between $\bar{I}$ and $(|E_g|/M)^{-2}$
\begin{equation}\label{UREqIEg}
    \bar{I} = 5.0757+ 2.2503\sqrt{x} + 0.4975x, 
\end{equation}
where we have introduced the new dimensionless variable $x= \bar{E}_g^{-2}$ with $\bar{E}_g = \vert E_g\vert/M$. Through the left panel of Fig.~\ref{FigLEgfEg}, we have presented the UR between the inverse $2^{\text{nd}}$ power of dimensionless GBE $\bar{E}_g^{-2}$ and the normalized MoI $\bar{I}$. Again, the lower panel of that figure displays the relative fractional difference between the numerical results and the fitting formula, with maximum deviation at the $2\%$ level under the variation of EoS parameter $\alpha$. Furthermore, when comparing our findings with those of Ref.~\cite{Jiang2019}, where the authors have obtained a linear UR of the form $\bar{I}= 0.1806x+ 9.314$, we observe that the $I-E_g^{-2}$ correlation for our DESs is substantially different from that obtained in the case of NSs and QSs. For completeness, we have also added the IQS case with a red dot-dashed curve.

Several precise URs involving the tidal deformability $\Lambda$ have been obtained, including the $I$–Love and $Q$–Love relations \cite{Yagi2013Science}, as well as correlations between the $f$-mode frequency/damping time and $\Lambda$ \cite{Wen2019}. The measurement of tidal deformability from gravitational wave (GW) signals has established it as a fundamental astrophysical property of NSs. In fact, the constraint on $\Lambda$ for a $1.4\, M_\odot$ compact star inferred from the GW170817 event is given by $\Lambda_{1.4} = 190^{+390}_{-120}$ at the $90\%$ confidence level \cite{Abbott2018PRL}, considering hadronic EoSs for the coalescing NSs. This motivates us to explore the connection between the tidal deformability and GBE for our DE stellar configurations. Interestingly, it has been observed that there exists an ideal linear UR between the negative fifth power of the dimensionless GBE and the tidal deformability for NSs and QSs \cite{Jiang2019}. We hence want to discuss the existence of such UR in the context of our proposed DESs. According to our calculations, this correlation can be approximated as
\begin{align}\label{UREqLEg}
    \log_{10}\Lambda =& -3.5215+ 45.1284\sqrt{y}+1.6460y  \nonumber  \\
    &- (2.5369\times 10^{-3})\sqrt{y^3} + (4.4715\times 10^{-6})y^2,
\end{align}
where $y = \bar{E}_g^{-5}$. The middle panel of Fig.~\ref{FigLEgfEg} illustrates the relation between the negative fifth power of the dimensionless GBE and $\Lambda$. The results obtained for DESs show a clear and substantial deviation from those reported in the literature for conventional compact stars. It is worth emphasizing that the $\Lambda-\bar{E}_g^{-5}$ correlation obtained for IQSs (red dot-dashed curve) is nearly indistinguishable from the relation reported by Jiang et al.~\cite{Jiang2019} for NSs (blue dashed line), while it exhibits a marked discrepancy with the empirical fit derived by the same authors for QSs described by confined-density-dependent-mass model (CDDM) EoSs (magenta dashed curve). Our study here shows that DESs obey $\Lambda- \bar{E}_g^{-5}$ URs very different from those already reported in the literature for NSs and QSs. Therefore, this UR also allows us to substantially distinguish DESs from compact stars made of hadronic matter and quark matter. In addition, if two of the three amounts $\{\Lambda, E_g, M\}$ are known precisely from the observations, then our $\Lambda- \bar{E}_g^{-5}$ UR allows us to determine the third macroscopic property of a compact star. This procedure is potentially useful when determining global quantities that cannot be measured observationally, and as we shall see below, it is also applied to other URs. The empirical formulas involving the GBE for IQSs are explicitly provided in Appendix \ref{URsforIQSs}.

Finally, we report that the inverse power of dimensionless GBE and $f$-mode pulsation frequency can be approximated by the following empirical formula
\begin{align}\label{UREqfEg}
    \Omega_f &= -0.01102+ \frac{0.43501}{\sqrt{x}} - \frac{0.54965}{x} + \frac{0.42268}{\sqrt{x^3}}  \nonumber  \\
    &-\frac{0.20546}{x^2} + \frac{6.4052\times 10^{-2}}{\sqrt{x^5}} - \frac{1.2725\times 10^{-2}}{x^3}   \nonumber  \\
    &+ \frac{1.5535\times 10^{-3}}{\sqrt{x^7}} - \frac{1.0602\times 10^{-4}}{x^4} + \frac{3.0919\times 10^{-6}}{\sqrt{x^9}} ,
\end{align}
where again the quantity $x$ has been defined by $x= \bar{E}_g^{-2}$. The UR (\ref{UREqfEg}) is illustrated in the right panel of Fig.~\ref{FigLEgfEg}. The fitted correlation is largely insensitive to variations in the parameter $\alpha$, displaying a maximum fractional deviation at the $\sim 3\%$ level. This demonstrates the robustness of the fit and its weak sensitivity to changes in $\alpha$, as shown in the lower right panel of the same figure. In addition, the $f-E_{g}^{-2}$ UR for IQSs differs markedly from that obtained using Chaplygin-type EoS, indicating that URs involving the GBE can be used to distinguish DESs from IQSs.

Remarkably, the use of our URs together with tidal deformability data allows one to place stringent constraints on key macroscopic properties of compact stars, such as the canonical radius and MoI for a $1.4\, M_{\odot}$ compact star. These quantities are particularly important, as they encode valuable information about the star's internal structure and the EoS of dense matter. The tidal deformability constraint $\Lambda_{1.4} = 190^{+390}_{-120}$, as reported by the LIGO-Virgo Collaboration from the GW170817 event~\cite{Abbott2018PRL}, is derived under the assumption that the merging objects were NSs. Nevertheless, since the present study focuses on DESs, we have to adopt the EoS-independent constraint $\Lambda_{1.4} \leq 800$ from the earlier work \cite{Abbott2017PRL}. Such a bound will allow us to constraint some properties of DESs from our empirical formulas involving $\Lambda$. In Fig.~\ref{FigICCL} (right panel) for the $C-\Lambda$ relation, Fig.~\ref{FigIL} for $I-\Lambda$ correlation, and Fig.~\ref{FigfL} for $f-\Lambda$ relationship, we have shown the tidal deformability constraint $\Lambda_{1.4} \leq 800$ by yellow shaded regions. By using our polynomial approximations, we can claim that the $C-\Lambda$ UR leads to $C_{1.4}\geq 0.176$, the $I-\Lambda$ UR allows us to obtain $\bar{I}_{1.4} \leq 14.966$, and the $f-\Lambda$ UR generates the canonical normalized $f$-mode frequency $\Omega_{f,1.4} \geq 0.092$. Consequently, in physical units, the canonical properties of a $1.4\, M_\odot$ compact star are equivalently written as $R_{1.4} \leq 11.738\, \rm km$, $I_{1.4}\leq 1.784\, \rm g\cdot cm^2$ and $f_{f,1.4} \geq 2.121\, \rm kHz$. Therefore, in the present study we have been able to impose certain limits on the canonical properties of DESs through their URs.

To achieve a more comprehensive and robust assessment of the universality of the proposed relations, we have included in Appendix \ref{extended} a systematic analysis of URs involving the tidal deformability $\Lambda$ under variations of the remaining free parameters of the MCG EoS. Specifically, we explore the cases in which the parameters $A$ and $B$ are varied independently. This extended study allows us to examine whether the approximate universality of the correlations persists beyond variations in $\alpha$ alone, and to assess their robustness across a broader region of the MCG parameter space.

\section{Final remarks}\label{section4}

In this work, we have constructed DESs described by the MCG model over a broad range of values of the free parameter $\alpha$, while consistently satisfying the causality condition. Under the assumption that the stellar configuration is completely made of a Chaplygin-type dark fluid, we computed several important properties, including the radius $R$, gravitational mass $M$, moment of inertia $I$, tidal deformability $\Lambda$, gravitational binding energy $E_g$ and $f$-mode nonradial pulsation frequency. Our $M-R$ analysis showed that decreasing $\alpha$ shifts the curves toward larger radii and higher masses, leading to a significant increase in the maximum supported mass. Remarkably, these properties are highly compatible with different observational measurements. On closer inspection, this is because the EoS becomes stiffer for small $\alpha$. As a consequence, for a fixed mass, both the MoI and the tidal deformability increase as $\alpha$ decreases, while the GBE and the fundamental nonradial oscillation frequency decrease. In addition, based on our numerical data, we derived a set of empirical functions that allow one to infer the canonical properties of a $1.4\,M_\odot$ compact star from tidal deformability measurements obtained through GW observations.

Our work advances the understanding of DESs by establishing correlations among their different global quantities. We provide several URs expressed through power-series expansions and systematically compare them with the corresponding relations for standard QSs and NSs. URs serve as powerful tools for deducing stellar characteristics that cannot be directly observed using existing detectors or telescopes. We have demonstrated that the $C-I-\Lambda-E_g-f$ URs generally hold for DESs, exhibiting maximum relative deviations at a very low percentage level between the numerical data and the fitted curves. The weak dependence of these relations on the MCG EoS parameter $\alpha$ indicates that they are effectively universal with respect to its variations.

Our findings revealed that the GBE encapsulates essential information about the internal mass distribution of compact stars and can therefore be employed as a fundamental quantity for constructing URs associated with the basic properties of isotropic DESs. In particular, we have established three GBE-based URs involving the normalized MoI, the tidal deformability, and the dimensionless $f$-mode oscillation frequency. We have shown that $I-E_{g}^{-2}$ and $\Lambda-E_{g}^{-5}$ correlations obtained for DESs do not coincide with the corresponding relations for normal NSs and QSs reported in the literature \cite{Jiang2019}. Motivated by Ref.~\cite{Pretel2024} and in order to enlarge the set of QS correlations available for comparison with our DES results, we have additionally derived new GBE-related URs for IQSs. Notably, we found that the $f-E_{g}^{-2}$ relation for DESs differs substantially from that predicted for IQSs, indicating that the GBE-based URs obtained here provide a clear means of discriminating DESs from NSs and QSs. Furthermore, we extended our analysis beyond variations in the parameter $\alpha$. As discussed in Appendix~\ref{extended}, we have also examined changes in the remaining parameters of the MCG EoS and found that all correlations involving the tidal deformability remain approximately universal under variations of $B$, with the sole exception of the $\Lambda-E_g^{-5}$ relation, which breaks down when the parameter $A$ is varied. Finally, by combining these URs with the observational constraint $\Lambda_{1.4}\leq 800$ inferred from the GW170817 event \cite{Abbott2017PRL}, we have placed meaningful constraints on the canonical properties of a $1.4\,M_{\odot}$ compact star.

Overall, this study highlights the relevance of understanding the fundamental macroscopic characteristics of hypothetical DESs governed by the MCG EoS. It is worth noting that the URs involving the MoI of DE stellar configurations were derived within the slow-rotation approximation and under the assumption of an isotropic perfect fluid. These limitations can be addressed in future investigations by incorporating higher-order rotational corrections to the frame-dragging angular velocity, allowing for anisotropic pressure components, and exploring additional physical effects such as strong magnetic fields, the possible presence of quark matter in the stellar core, and contributions from dark matter.

\begin{acknowledgments}
We thank the Editor and the anonymous reviewer for their insightful comments and constructive suggestions, which have significantly improved the clarity and physical soundness of this manuscript. JMZP acknowledges support from ``Fundação Carlos Chagas Filho de Amparo à Pesquisa do Estado do Rio de Janeiro'' -- FAPERJ, Process SEI-260003/000308/2024.
\end{acknowledgments}

\begin{figure*}
\includegraphics[width=8.2cm]{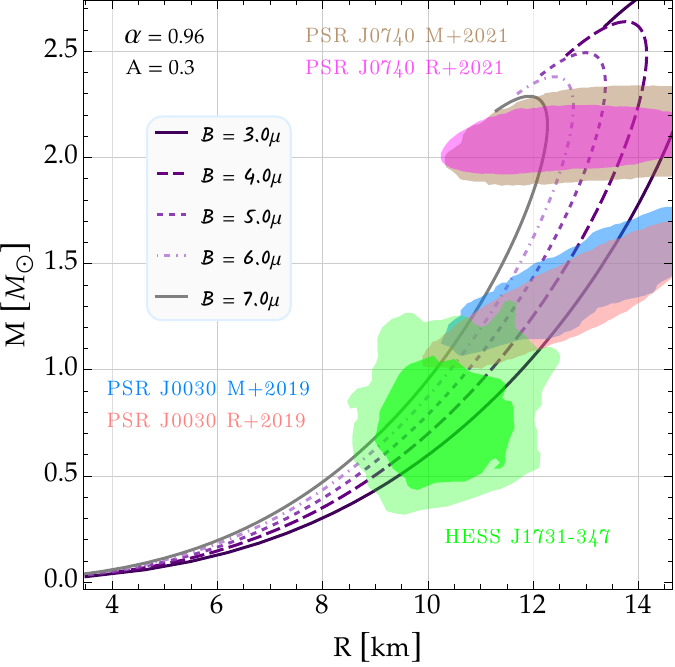}
\hspace{2mm}
\includegraphics[width=8.2cm]{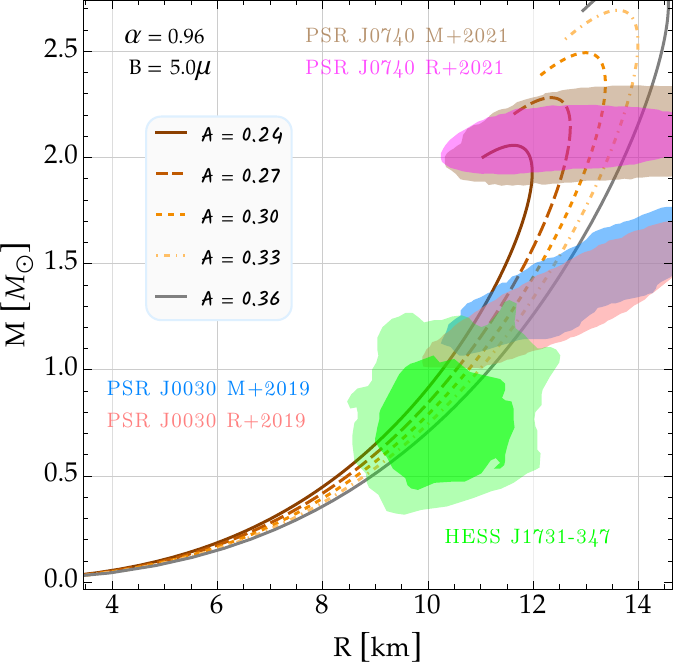}
    \caption{Mass-radius predictions for the DE stellar models I (left) and II (right), where the colored regions represent the same as in Fig.~\ref{FigMR}. }
    \label{FigMRAandBvarying}%
\end{figure*}

\appendix
\section{Empirical formulas for IQSs}\label{URsforIQSs}

For comparison, and following Ref.~\cite{Pretel2024}, the plots in Fig.~\ref{FigLEgfEg} also include the empirical relations obtained for isotropic IQSs, shown as red dot-dashed curves. In that study, URs involving $I-{\rm Love}-f-C$ were reported, although the GBE was not considered. To enable a more meaningful comparison with our DES results, we therefore derived the corresponding correlations involving dimensionless $E_g$ for IQSs. As in Ref.~\cite{Pretel2024}, we considered several interacting quark-matter EoSs characterized by $\tilde{\lambda} = 0, 1, \infty$, where $\tilde{\lambda} = \lambda^2/(4B_{\rm eff})$, $B_{\rm eff}$ denotes the effective bag constant, and $\lambda$ provides a relative measure of strong-interaction effects \cite{Zhang2021}. The correlations $\bar{I}-\bar{E}_g^{-2}$, $\Lambda-\bar{E}_g^{-5}$ and $f-\bar{E}_g^{-2}$ for isotropic IQSs are given respectively by
\begin{equation}
    \bar{I} = 3.0726+ 0.2351\sqrt{x}+ 0.1983x- (2.0230\times 10^{-3})\sqrt{x^3} ,
\end{equation}
\begin{align}
    \log_{10}\Lambda =&\ 1.6529- 0.1874\sqrt{y}+ (3.6108\times 10^{-2})y  \nonumber  \\
    &+ (2.4648\times 10^{-6})\sqrt{y^3} ,
\end{align}
and 
\begin{align}
    \Omega_f =& -0.01943+ \frac{0.94642}{\sqrt{x}} - \frac{0.63960}{x} - \frac{2.5891}{\sqrt{x^3}}  \nonumber  \\
    &+ \frac{6.3253}{x^2}- \frac{5.2908}{\sqrt{x^5}} , 
\end{align}
where, as defined before, $x= \bar{E}_g^{-2}$ and $y= \bar{E}_g^{-5}$. Beyond the analysis of DESs, this work presents—for the first time to the best of our knowledge—empirical URs involving the GBE for IQSs. These results extend the existing framework of URs and provide a robust benchmark for future theoretical and observational studies aimed at probing strong-interaction effects in compact stars.

\begin{figure*}
\includegraphics[width=8.4cm]{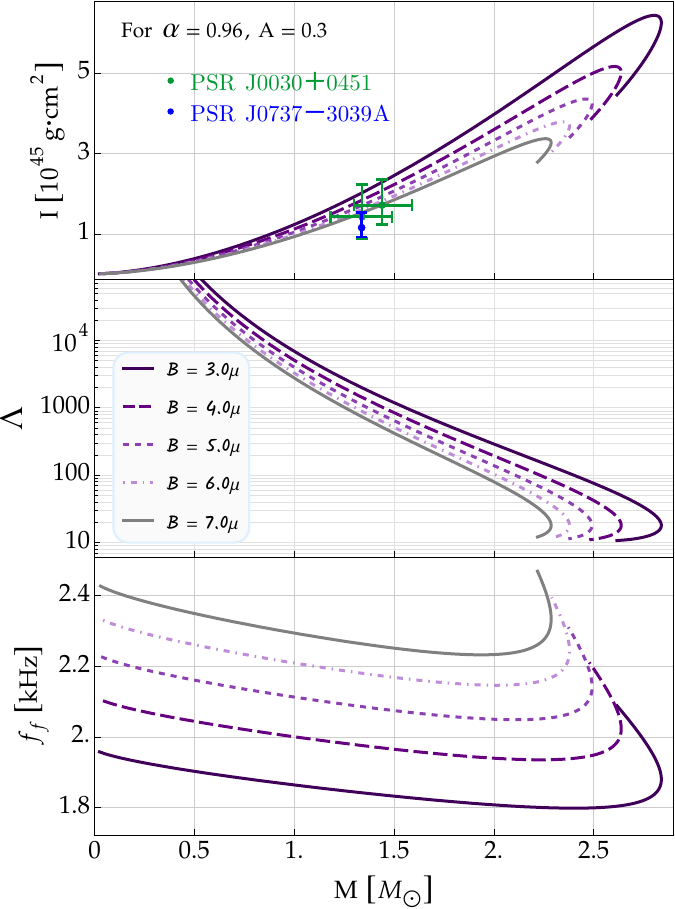}
\hspace{2mm}
\includegraphics[width=8.4cm]{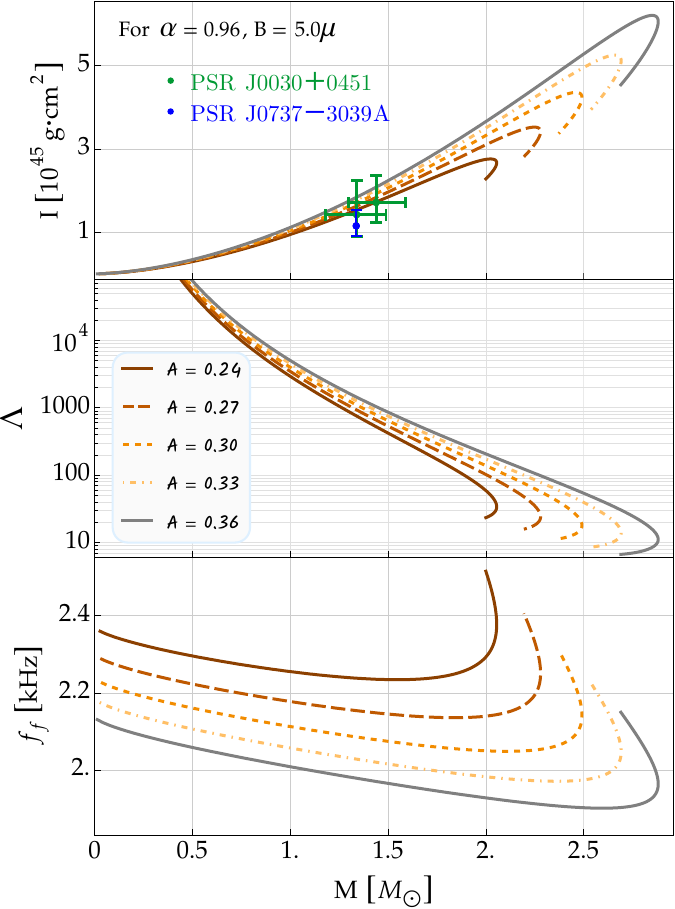}
    \caption{MoI (upper), tidal deformability (middle) and fundamental nonradial pulsation frequency (lower) as functions of the gravitational mass $M$ for the stellar configurations presented in Fig.~\ref{FigMRAandBvarying}. }
    \label{FigILfAandBvarying}%
\end{figure*}

\begin{table}
\begin{minipage}[b]{86mm}
\caption{\label{tableCaseI} Fitting parameters of the $C-\Lambda$ \eqref{UREqCLambda}, $I-\Lambda$ \eqref{UREqILambda} and $f-\Lambda$ \eqref{UREqfLambda} URs for case I. }
\begin{ruledtabular}
\begin{tabular}{c|c|c}
$C-\Lambda$ UR  &  $I-\Lambda$ UR  &  $f-\Lambda$ UR  \\
\hline
  $b_0 [10^{-1}]= 2.9134$  &  $c_0 [10^{-1}]= 6.3909$  &  $h_0 [10^{-1}]= 1.2841$  \\
  $b_1 [10^{-2}]= 5.9991$  &  $c_1 [10^{-2}]= 7.4197$  &  $h_1 [10^{-1}]= 1.3024$  \\
  $b_2 [10^{-2}]= -7.4823$  &  $c_2 [10^{-2}]= 4.5660$  &  $h_2 [10^{-1}]= -1.2231$  \\
  $b_3 [10^{-2}]= 1.9589$  &  $c_3 [10^{-3}]= -2.7641$  &  $h_3 [10^{-2}]= 4.1402$  \\
  $b_4 [10^{-3}]= -2.2171$  &  $c_4 [10^{-5}]= 5.0786$  &  $h_4 [10^{-3}]= -7.3471$  \\
  $b_5 [10^{-5}]= 9.4673$  &  $--$  &  $h_5 [10^{-4}]= 6.8183$  \\
  $--$  &  $--$  &  $h_6 [10^{-5}]= -2.6010$
\end{tabular}
\end{ruledtabular}
\end{minipage}
\end{table}

\begin{table}
\begin{minipage}[b]{86mm}
\caption{\label{tableCaseII} Fitting coefficients of the $C-\Lambda$ \eqref{UREqCLambda}, $I-\Lambda$ \eqref{UREqILambda} and $f-\Lambda$ \eqref{UREqfLambda} URs for case II. }
\begin{ruledtabular}
\begin{tabular}{c|c|c}
$C-\Lambda$ UR  &  $I-\Lambda$ UR  &  $f-\Lambda$ UR  \\
\hline
  $b_0 [10^{-1}]= 3.5247$  &  $c_0 [10^{-1}]= 6.4692$  &  $h_0 [10^{-1}]= 2.1801$  \\
  $b_1 [10^{-2}]= -4.4356$  &  $c_1 [10^{-2}]= 6.3226$  &  $h_1 [10^{-2}]= -5.6174$  \\
  $b_2 [10^{-2}]= -1.3171$  &  $c_2 [10^{-2}]= 5.0409$  &  $h_2 [10^{-2}]= 2.4260$  \\
  $b_3 [10^{-3}]= 3.1587$  &  $c_3 [10^{-3}]= -3.5719$  &  $h_3 [10^{-2}]= -1.5527$  \\
  $b_4 [10^{-4}]= -1.9711$  &  $c_4 [10^{-5}]= 9.7883$  &  $h_4 [10^{-3}]= 4.3292$  \\
  $b_5 [10^{-6}]= 1.6122$  &  $--$  &  $h_5 [10^{-4}]= -5.2895$  \\
  $--$  &  $--$  &  $h_6 [10^{-5}]= 2.3961$
\end{tabular}
\end{ruledtabular}
\end{minipage}
\end{table}

\begin{table*}
\begin{minipage}[b]{120mm}
\caption{\label{table4} Theoretical bounds for the canonical global properties of a $1.4\, M_\odot$ compact star from the tidal deformability constraint given by the GW170817 signal \cite{Abbott2017PRL}.  }
\begin{ruledtabular}
\begin{tabular}{c|c|c|c|c}
Case  &  $R\, [\rm km]$  &  $I\, [10^{45}\, \rm g\cdot cm^2]$  &  $f_f =\nu_f/2\pi\, [\rm kHz]$  &  $E_g\, [M_\odot]$  \\
\hline
  I  &  $R_{1.4}\leq 11.740$  &  $I_{1.4} \leq 1.785$  &  $f_{f,1.4} \geq 2.118$  &  $E_{g, 1.4}\leq -0.515$  \\
  II  &  $R_{1.4}\leq 11.734$  &  $I_{1.4} \leq 1.784$  &  $f_{f,1.4} \geq 2.127$  &  $--$
\end{tabular}
\end{ruledtabular}
\end{minipage}
\end{table*}

\section{Extensive Parameter Scan}\label{extended}

Since our stellar models include three free parameters, i.e., $\{A, B, \alpha\}$, here we will explore a more general parameter scan by considering a sizable parameter space whose $M-R$ predictions are compatible with observational measurements. Our results above have focused on varying $\alpha$, but have kept both $A$ and $B$ fixed, so to make our study more general and robust for DES URs, we will adopt two more scenarios:
\begin{itemize}
    \item[$\star$] \textbf{Case I:} $\alpha= 0.96$ and $A= 0.3$, but varying $B$ from $3.0$ to $7.0 \mu$, where $\mu= 10^{-20}\, \rm m^{-3.92}$.
    \item[$\star$] \textbf{Case II:} $\alpha= 0.96$ and $B= 5.0\mu$, with $A$ varying in the range $A\in [0.24, 0.36]$.
\end{itemize}

The various global properties $\{ R,M,I,\Lambda,f_f \}$ for such configurations made from the MCG are exhibited in figures \ref{FigMRAandBvarying} and \ref{FigILfAandBvarying}. For case I, one can observe that the effect of an increasing $B$ is a noticeable decrease in the maximum mass values. Note further that both $I$ and $\Lambda$ decrease with increasing $B$ for a given mass value $M$, while the opposite occurs for the $f$-mode frequency. Meanwhile, for case II, increasingly larger values of $A$ lead to an increase in maximum mass and MoI, but $f_f$ decreases.

\begin{figure*}
\includegraphics[width=8.4cm]{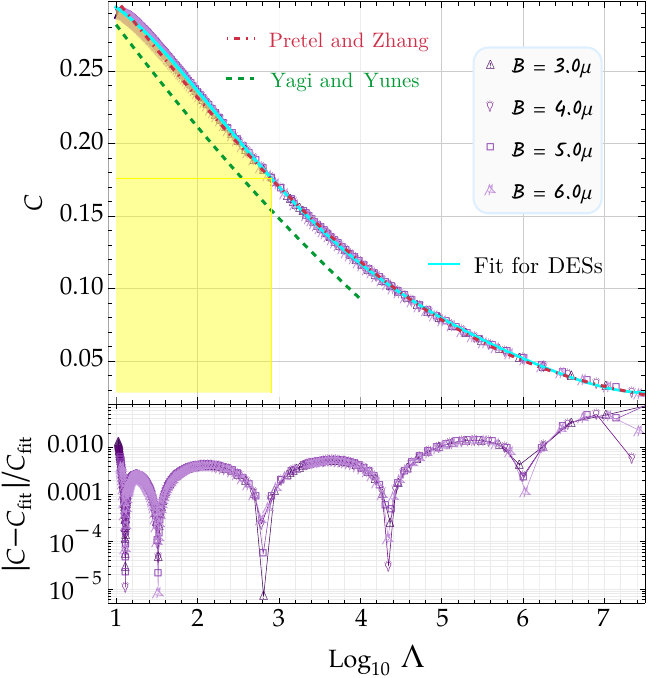}
\hspace{2mm}
\includegraphics[width=8.38cm]{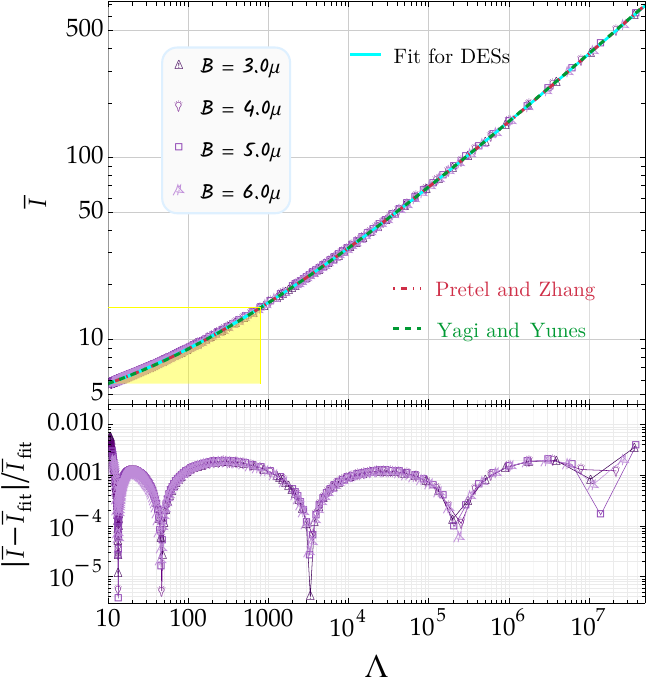}
\includegraphics[width=8.4cm]{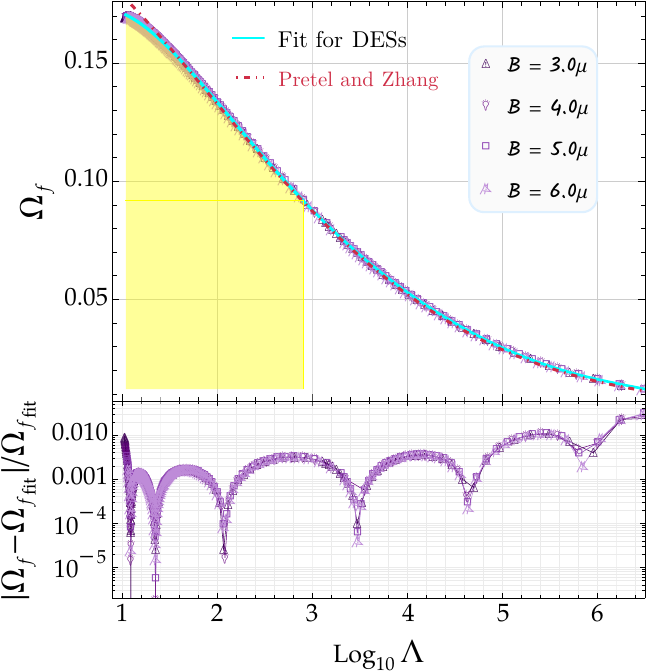}
\hspace{2mm}
\includegraphics[width=8.4cm]{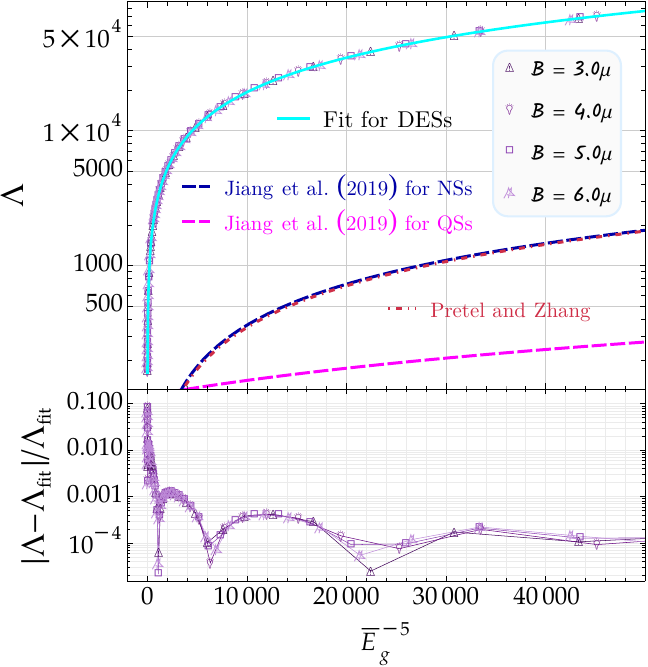}
    \caption{ $C-\Lambda$ (top left), $I-\Lambda$ (top right), $f-\Lambda$ (bottom left), and $\Lambda-E_g^{-5}$ (bottom right) URs corresponding to case I, where we have varied $B$ but kept both $\alpha$ and $A$ fixed. The power-series expansions are represented by the cyan curves, while the corresponding fit residuals are displayed in the lower panels of each plot. The red, green, blue, and magenta curves, as well as the yellow shaded regions, represent the same features as in the scenario where $\alpha$ is varied. }
    \label{FigURsCaseI}%
\end{figure*}

\begin{figure*}
\includegraphics[width=8.4cm]{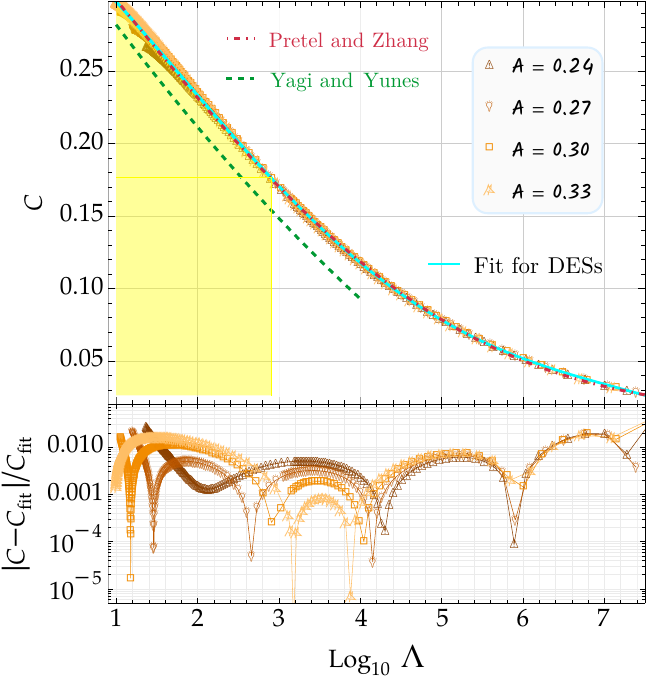}
\hspace{2mm}
\includegraphics[width=8.38cm]{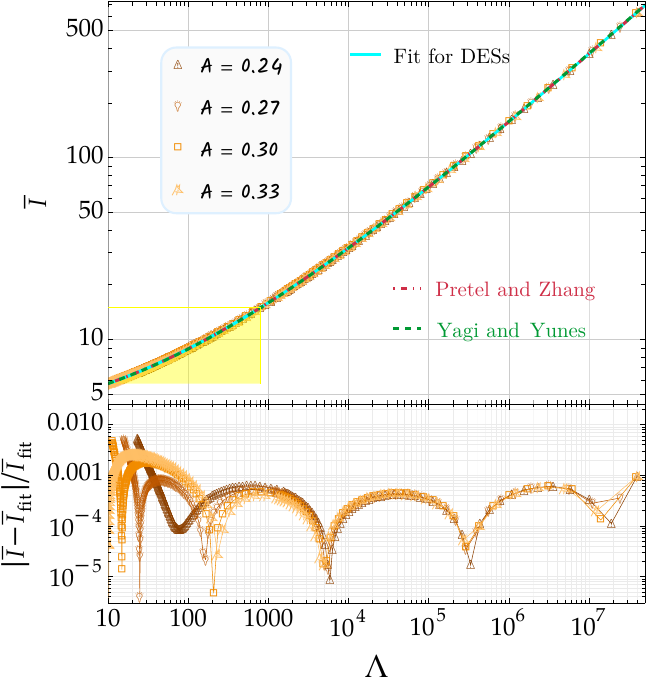}
\includegraphics[width=8.4cm]{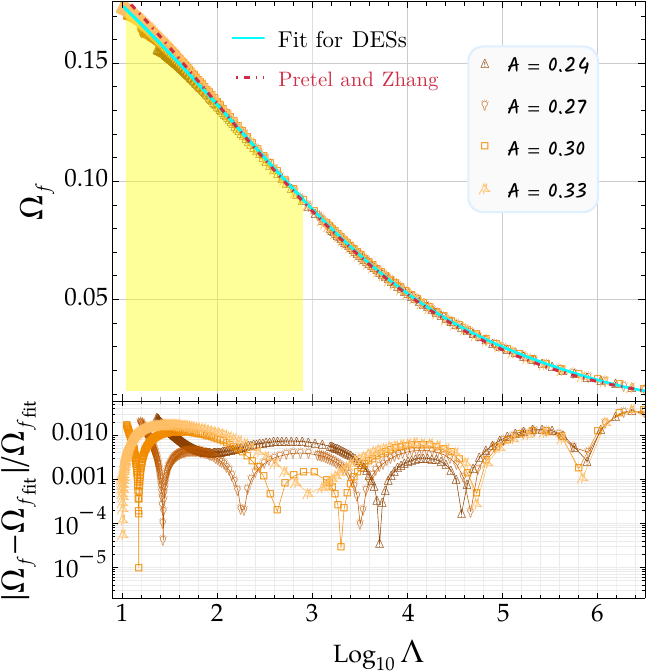}
\hspace{2mm}
\includegraphics[width=8.4cm]{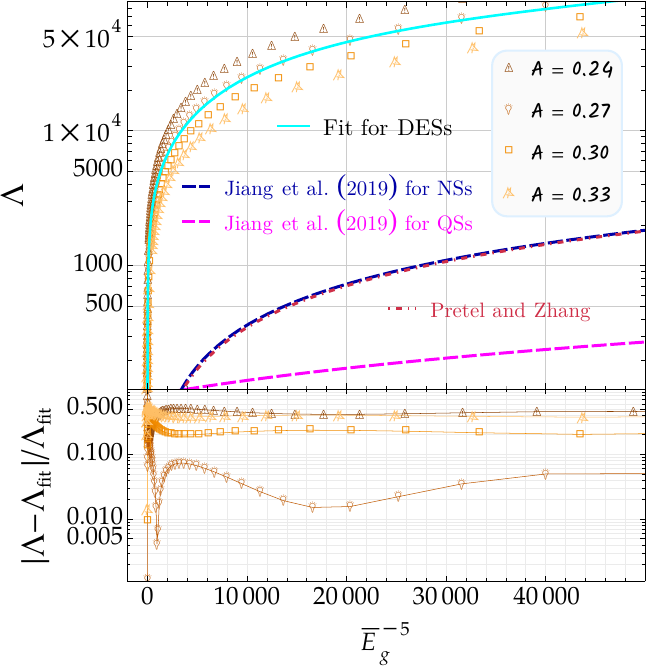}
    \caption{Various relations involving $\Lambda$ for case II, as in Fig.~\ref{FigURsCaseI}, where we considered four distinct values of the MCG parameter $A$ with the cyan curves corresponding to our fitting formulas. A notable feature is that universality is violated in the $\Lambda-E_g^{-5}$ relation when $A$ is varied, as this relation exhibits very large fractional errors. }
    \label{FigURsCaseII}%
\end{figure*}

To avoid repeating all the URs obtained above, we will focus here on the correlations involving $\Lambda$ because these are the ones that allow us to obtain other unmeasured stellar quantities. Figures \ref{FigURsCaseI} and \ref{FigURsCaseII} show the different URs for cases I and II, where their fitting coefficients are given in tables \ref{tableCaseI} and \ref{tableCaseII}, respectively. For case I, we observe that variations of the parameter $B$ (with both $\alpha$ and $A$ fixed) lead to fractional differences approximately below $9\%$ between the data and the empirical formulas indicating that universality is maintained for variations of $B$ to $\sim 9\%$ level. For this case, the $\Lambda-E_g^{-5}$ correlation can be represented by the following fit function
\begin{align}\label{UREqLEgCaseI}
    \log_{10}\Lambda =& -15.2610+ 47.3491\sqrt{y}+ 1.6547y  \nonumber  \\
    &- (2.7777\times 10^{-3})\sqrt{y^3} + (7.7476\times 10^{-6})y^2  \nonumber  \\
    &- (9.0491\times 10^{-9})\sqrt{y^5} ,
\end{align}
which is shown by the cyan curve in the bottom right plot of Fig.~\ref{FigURsCaseI}.

On the other hand, from Fig.~\ref{FigURsCaseII}, one notices that the $C-\Lambda$, $I-\Lambda$ and $f-\Lambda$ relations are insensitive to variations of $A$ (with both $\alpha$ and $B$ fixed) to $\sim 4\%$ level. Nevertheless, the $\Lambda-E_g^{-5}$ relation exhibits a strong dependence on the parameter $A$ and is therefore not universal since in such case the fractional deviations can be up to $80\%$. While the $\Lambda-E_g^{-5}$ relation is universal at the $\sim 8\%$ level for variations in $\alpha$ and $B$, it breaks down when parameter $A$ is varied, leading to deviations of up to $80\%$. For this reason, in table \ref{table4} we have only provided the bound of $E_{g,1.4}$ for case I. It is important to emphasize that we have adopted variations of $A$ and $B$ such that the range of gravitational masses is astrophysically relevant (see Fig.~\ref{FigMRAandBvarying}), so that relations with fractional errors $\lesssim 10\%$ are acceptable as universal, while above that percentage the universality could be broken. 

Similar to the predictions we obtained from the URs by varying the free parameter $\alpha$ for a $1.4\, M_\odot$ compact star from the observational constraint for the GW170817 signal, we can also infer here some limits for the radius, moment of inertia, fundamental oscillation frequency and gravitational binding energy taking into account the two extra cases; see table \ref{table4}. The theoretical bound obtained for both the radius and the fundamental nonradial pulsation frequency changes in the second decimal place; however, the moment of inertia is almost the same in both cases (with a small difference only in the third decimal place). These upper-limit values of $I_{1.4}$, which are closer than those obtained for other canonical properties, arise from the particularly small fractional error associated with the $I-\Lambda$ UR. Indeed, the error remains below $\sim 0.6\%$, indicating universality at the sub-percent level.


\newpage
\begin{thebibliography}{99}%
\makeatletter
\providecommand \@ifxundefined [1]{%
 \@ifx{#1\undefined}
}%
\providecommand \@ifnum [1]{%
 \ifnum #1\expandafter \@firstoftwo
 \else \expandafter \@secondoftwo
 \fi
}%
\providecommand \@ifx [1]{%
 \ifx #1\expandafter \@firstoftwo
 \else \expandafter \@secondoftwo
 \fi
}%
\providecommand \natexlab [1]{#1}%
\providecommand \enquote  [1]{``#1''}%
\providecommand \bibnamefont  [1]{#1}%
\providecommand \bibfnamefont [1]{#1}%
\providecommand \citenamefont [1]{#1}%
\providecommand \href@noop [0]{\@secondoftwo}%
\providecommand \href [0]{\begingroup \@sanitize@url \@href}%
\providecommand \@href[1]{\@@startlink{#1}\@@href}%
\providecommand \@@href[1]{\endgroup#1\@@endlink}%
\providecommand \@sanitize@url [0]{\catcode `\\12\catcode `\$12\catcode `\&12\catcode `\#12\catcode `\^12\catcode `\_12\catcode `\%12\relax}%
\providecommand \@@startlink[1]{}%
\providecommand \@@endlink[0]{}%
\providecommand \url  [0]{\begingroup\@sanitize@url \@url }%
\providecommand \@url [1]{\endgroup\@href {#1}{\urlprefix }}%
\providecommand \urlprefix  [0]{URL }%
\providecommand \Eprint [0]{\href }%
\providecommand \doibase [0]{http://dx.doi.org/}%
\providecommand \selectlanguage [0]{\@gobble}%
\providecommand \bibinfo  [0]{\@secondoftwo}%
\providecommand \bibfield  [0]{\@secondoftwo}%
\providecommand \translation [1]{[#1]}%
\providecommand \BibitemOpen [0]{}%
\providecommand \bibitemStop [0]{}%
\providecommand \bibitemNoStop [0]{.\EOS\space}%
\providecommand \EOS [0]{\spacefactor3000\relax}%
\providecommand \BibitemShut  [1]{\csname bibitem#1\endcsname}%
\let\auto@bib@innerbib\@empty
\bibitem [{\citenamefont {Perlmutter}\ \emph {et~al.}(1998)\citenamefont {Perlmutter} \emph {et~al.}}]{perlmutter1998discovery}%
  \BibitemOpen
  \bibfield  {author} {\bibinfo {author} {\bibfnamefont {S.}~\bibnamefont {Perlmutter}} \emph {et~al.},\ }\href {\doibase doi.org/10.1038/34124} {\bibfield  {journal} {\bibinfo  {journal} {Nature}\ }\textbf {\bibinfo {volume} {391}},\ \bibinfo {pages} {51} (\bibinfo {year} {1998})}\BibitemShut {NoStop}%
\bibitem [{\citenamefont {Riess}\ \emph {et~al.}(1998)\citenamefont {Riess} \emph {et~al.}}]{riess1998observational}%
  \BibitemOpen
  \bibfield  {author} {\bibinfo {author} {\bibfnamefont {A.~G.}\ \bibnamefont {Riess}} \emph {et~al.},\ }\href {\doibase 10.1086/300499} {\bibfield  {journal} {\bibinfo  {journal} {Astron. J.}\ }\textbf {\bibinfo {volume} {116}},\ \bibinfo {pages} {1009} (\bibinfo {year} {1998})}\BibitemShut {NoStop}%
\bibitem [{\citenamefont {Hicken}\ \emph {et~al.}(2009)\citenamefont {Hicken} \emph {et~al.}}]{hicken2009improved}%
  \BibitemOpen
  \bibfield  {author} {\bibinfo {author} {\bibfnamefont {M.}~\bibnamefont {Hicken}} \emph {et~al.},\ }\href {\doibase 10.1088/0004-637X/700/2/1097} {\bibfield  {journal} {\bibinfo  {journal} {Astrophys. J.}\ }\textbf {\bibinfo {volume} {700}},\ \bibinfo {pages} {1097} (\bibinfo {year} {2009})}\BibitemShut {NoStop}%
\bibitem [{\citenamefont {Seikel}\ and\ \citenamefont {Schwarz}(2009)}]{seikel2009model}%
  \BibitemOpen
  \bibfield  {author} {\bibinfo {author} {\bibfnamefont {M.}~\bibnamefont {Seikel}}\ and\ \bibinfo {author} {\bibfnamefont {D.~J.}\ \bibnamefont {Schwarz}},\ }\href {\doibase 10.1088/1475-7516/2009/02/024} {\bibfield  {journal} {\bibinfo  {journal} {JCAP}\ }\textbf {\bibinfo {volume} {02}},\ \bibinfo {pages} {024} (\bibinfo {year} {2009})}\BibitemShut {NoStop}%
\bibitem [{\citenamefont {Aghanim}\ \emph {et~al.}(2020)\citenamefont {Aghanim} \emph {et~al.}}]{aghanim2020planck}%
  \BibitemOpen
  \bibfield  {author} {\bibinfo {author} {\bibfnamefont {N.}~\bibnamefont {Aghanim}} \emph {et~al.},\ }\href {\doibase 10.1051/0004-6361/201833910} {\bibfield  {journal} {\bibinfo  {journal} {Astron. Astrophys}\ }\textbf {\bibinfo {volume} {641}},\ \bibinfo {pages} {A6} (\bibinfo {year} {2020})}\BibitemShut {NoStop}%
\bibitem [{\citenamefont {Kamenshchik}\ \emph {et~al.}(2001)\citenamefont {Kamenshchik}, \citenamefont {Moschella},\ and\ \citenamefont {Pasquier}}]{Kamenshchik2001}%
  \BibitemOpen
  \bibfield  {author} {\bibinfo {author} {\bibfnamefont {A.}~\bibnamefont {Kamenshchik}}, \bibinfo {author} {\bibfnamefont {U.}~\bibnamefont {Moschella}}, \ and\ \bibinfo {author} {\bibfnamefont {V.}~\bibnamefont {Pasquier}},\ }\href {\doibase https://doi.org/10.1016/S0370-2693(01)00571-8} {\bibfield  {journal} {\bibinfo  {journal} {Phys. Lett. B}\ }\textbf {\bibinfo {volume} {511}},\ \bibinfo {pages} {265} (\bibinfo {year} {2001})}\BibitemShut {NoStop}%
\bibitem [{\citenamefont {Bili{\'c}}\ \emph {et~al.}(2002)\citenamefont {Bili{\'c}}, \citenamefont {Tupper},\ and\ \citenamefont {Viollier}}]{bilic2002unification}%
  \BibitemOpen
  \bibfield  {author} {\bibinfo {author} {\bibfnamefont {N.}~\bibnamefont {Bili{\'c}}}, \bibinfo {author} {\bibfnamefont {G.~B.}\ \bibnamefont {Tupper}}, \ and\ \bibinfo {author} {\bibfnamefont {R.~D.}\ \bibnamefont {Viollier}},\ }\href {\doibase 10.1016/S0370-2693(02)01716-1} {\bibfield  {journal} {\bibinfo  {journal} {Phys. Lett. B}\ }\textbf {\bibinfo {volume} {535}},\ \bibinfo {pages} {17} (\bibinfo {year} {2002})}\BibitemShut {NoStop}%
\bibitem [{\citenamefont {Makler}\ \emph {et~al.}(2003)\citenamefont {Makler}, \citenamefont {{de Oliveira}},\ and\ \citenamefont {Waga}}]{Makler2003}%
  \BibitemOpen
  \bibfield  {author} {\bibinfo {author} {\bibfnamefont {M.}~\bibnamefont {Makler}}, \bibinfo {author} {\bibfnamefont {S.~Q.}\ \bibnamefont {{de Oliveira}}}, \ and\ \bibinfo {author} {\bibfnamefont {I.}~\bibnamefont {Waga}},\ }\href {\doibase https://doi.org/10.1016/S0370-2693(03)00038-8} {\bibfield  {journal} {\bibinfo  {journal} {Phys. Lett. B}\ }\textbf {\bibinfo {volume} {555}},\ \bibinfo {pages} {1} (\bibinfo {year} {2003})}\BibitemShut {NoStop}%
\bibitem [{\citenamefont {Debnath}\ \emph {et~al.}(2004)\citenamefont {Debnath}, \citenamefont {Banerjee},\ and\ \citenamefont {Chakraborty}}]{debnath2004role}%
  \BibitemOpen
  \bibfield  {author} {\bibinfo {author} {\bibfnamefont {U.}~\bibnamefont {Debnath}}, \bibinfo {author} {\bibfnamefont {A.}~\bibnamefont {Banerjee}}, \ and\ \bibinfo {author} {\bibfnamefont {S.}~\bibnamefont {Chakraborty}},\ }\href {\doibase 10.1088/0264-9381/21/23/019} {\bibfield  {journal} {\bibinfo  {journal} {Class. Quantum Grav.}\ }\textbf {\bibinfo {volume} {21}},\ \bibinfo {pages} {5609} (\bibinfo {year} {2004})}\BibitemShut {NoStop}%
\bibitem [{\citenamefont {Pourhassan}\ and\ \citenamefont {Kahya}(2014)}]{pourhassan2014extended}%
  \BibitemOpen
  \bibfield  {author} {\bibinfo {author} {\bibfnamefont {B.}~\bibnamefont {Pourhassan}}\ and\ \bibinfo {author} {\bibfnamefont {E.}~\bibnamefont {Kahya}},\ }\href {\doibase 10.1016/j.rinp.2014.05.007} {\bibfield  {journal} {\bibinfo  {journal} {Results Phys.}\ }\textbf {\bibinfo {volume} {4}},\ \bibinfo {pages} {101} (\bibinfo {year} {2014})}\BibitemShut {NoStop}%
\bibitem [{\citenamefont {Zheng}\ \emph {et~al.}(2022)\citenamefont {Zheng} \emph {et~al.}}]{Zheng2022}%
  \BibitemOpen
  \bibfield  {author} {\bibinfo {author} {\bibfnamefont {J.}~\bibnamefont {Zheng}} \emph {et~al.},\ }\href {\doibase https://doi.org/10.1140/epjc/s10052-022-10517-4} {\bibfield  {journal} {\bibinfo  {journal} {Eur. Phys. J. C}\ }\textbf {\bibinfo {volume} {82}},\ \bibinfo {pages} {582} (\bibinfo {year} {2022})}\BibitemShut {NoStop}%
\bibitem [{\citenamefont {Bento}\ \emph {et~al.}(2003)\citenamefont {Bento}, \citenamefont {Bertolami},\ and\ \citenamefont {Sen}}]{Bento2003}%
  \BibitemOpen
  \bibfield  {author} {\bibinfo {author} {\bibfnamefont {M.}~\bibnamefont {Bento}}, \bibinfo {author} {\bibfnamefont {O.}~\bibnamefont {Bertolami}}, \ and\ \bibinfo {author} {\bibfnamefont {A.}~\bibnamefont {Sen}},\ }\href {\doibase https://doi.org/10.1016/j.physletb.2003.08.017} {\bibfield  {journal} {\bibinfo  {journal} {Phys. Lett. B}\ }\textbf {\bibinfo {volume} {575}},\ \bibinfo {pages} {172} (\bibinfo {year} {2003})}\BibitemShut {NoStop}%
\bibitem [{\citenamefont {Lu}\ \emph {et~al.}(2009)\citenamefont {Lu}, \citenamefont {Gui},\ and\ \citenamefont {Xu}}]{Lu2022}%
  \BibitemOpen
  \bibfield  {author} {\bibinfo {author} {\bibfnamefont {J.}~\bibnamefont {Lu}}, \bibinfo {author} {\bibfnamefont {Y.}~\bibnamefont {Gui}}, \ and\ \bibinfo {author} {\bibfnamefont {L.}~\bibnamefont {Xu}},\ }\href {\doibase https://doi.org/10.1140/epjc/s10052-009-1118-8} {\bibfield  {journal} {\bibinfo  {journal} {Eur. Phys. J. C}\ }\textbf {\bibinfo {volume} {63}},\ \bibinfo {pages} {349} (\bibinfo {year} {2009})}\BibitemShut {NoStop}%
\bibitem [{\citenamefont {Lian}\ \emph {et~al.}(2021)\citenamefont {Lian} \emph {et~al.}}]{Lian2021}%
  \BibitemOpen
  \bibfield  {author} {\bibinfo {author} {\bibfnamefont {Y.}~\bibnamefont {Lian}} \emph {et~al.},\ }\href {\doibase 10.1093/mnras/stab1373} {\bibfield  {journal} {\bibinfo  {journal} {MNRAS}\ }\textbf {\bibinfo {volume} {505}},\ \bibinfo {pages} {2111} (\bibinfo {year} {2021})}\BibitemShut {NoStop}%
\bibitem [{\citenamefont {Bento}\ \emph {et~al.}(2002)\citenamefont {Bento}, \citenamefont {Bertolami},\ and\ \citenamefont {Sen}}]{bento2002generalized}%
  \BibitemOpen
  \bibfield  {author} {\bibinfo {author} {\bibfnamefont {M.~C.}\ \bibnamefont {Bento}}, \bibinfo {author} {\bibfnamefont {O.}~\bibnamefont {Bertolami}}, \ and\ \bibinfo {author} {\bibfnamefont {A.~A.}\ \bibnamefont {Sen}},\ }\href {\doibase 10.1103/PhysRevD.66.043507} {\bibfield  {journal} {\bibinfo  {journal} {Phys. Rev. D}\ }\textbf {\bibinfo {volume} {66}},\ \bibinfo {pages} {043507} (\bibinfo {year} {2002})}\BibitemShut {NoStop}%
\bibitem [{\citenamefont {Panotopoulos}\ \emph {et~al.}(2021)\citenamefont {Panotopoulos}, \citenamefont {Rinc{\'o}n},\ and\ \citenamefont {Lopes}}]{panotopoulos2021slowly}%
  \BibitemOpen
  \bibfield  {author} {\bibinfo {author} {\bibfnamefont {G.}~\bibnamefont {Panotopoulos}}, \bibinfo {author} {\bibfnamefont {{\'A}.}~\bibnamefont {Rinc{\'o}n}}, \ and\ \bibinfo {author} {\bibfnamefont {I.}~\bibnamefont {Lopes}},\ }\href {\doibase 10.1016/j.dark.2021.100885} {\bibfield  {journal} {\bibinfo  {journal} {Phys. Dark Univ.}\ }\textbf {\bibinfo {volume} {34}},\ \bibinfo {pages} {100885} (\bibinfo {year} {2021})}\BibitemShut {NoStop}%
\bibitem [{\citenamefont {Pretel}(2023)}]{pretel2023radial}%
  \BibitemOpen
  \bibfield  {author} {\bibinfo {author} {\bibfnamefont {J.~M.~Z.}\ \bibnamefont {Pretel}},\ }\href {\doibase 10.1140/epjc/s10052-023-11198-3} {\bibfield  {journal} {\bibinfo  {journal} {Eur. Phys. J. C}\ }\textbf {\bibinfo {volume} {83}},\ \bibinfo {pages} {26} (\bibinfo {year} {2023})}\BibitemShut {NoStop}%
\bibitem [{\citenamefont {Bhattacharjee}\ and\ \citenamefont {Chattopadhyay}(2024)}]{Bhattacharjee2024}%
  \BibitemOpen
  \bibfield  {author} {\bibinfo {author} {\bibfnamefont {D.}~\bibnamefont {Bhattacharjee}}\ and\ \bibinfo {author} {\bibfnamefont {P.~K.}\ \bibnamefont {Chattopadhyay}},\ }\href {\doibase https://doi.org/10.1140/epjc/s10052-024-12449-7} {\bibfield  {journal} {\bibinfo  {journal} {Eur. Phys. J. C}\ }\textbf {\bibinfo {volume} {84}},\ \bibinfo {pages} {77} (\bibinfo {year} {2024})}\BibitemShut {NoStop}%
\bibitem [{\citenamefont {Jyothilakshmi}\ \emph {et~al.}(2024)\citenamefont {Jyothilakshmi}, \citenamefont {Naik},\ and\ \citenamefont {Sreekanth}}]{Jyothilakshmi2024}%
  \BibitemOpen
  \bibfield  {author} {\bibinfo {author} {\bibfnamefont {O.~P.}\ \bibnamefont {Jyothilakshmi}}, \bibinfo {author} {\bibfnamefont {L.~J.}\ \bibnamefont {Naik}}, \ and\ \bibinfo {author} {\bibfnamefont {V.}~\bibnamefont {Sreekanth}},\ }\href {\doibase https://doi.org/10.1140/epjc/s10052-024-12776-9} {\bibfield  {journal} {\bibinfo  {journal} {Eur. Phys. J. C}\ }\textbf {\bibinfo {volume} {84}},\ \bibinfo {pages} {427} (\bibinfo {year} {2024})}\BibitemShut {NoStop}%
\bibitem [{\citenamefont {Banerjee}\ \emph {et~al.}(2025)\citenamefont {Banerjee} \emph {et~al.}}]{banerjee2025EPJC}%
  \BibitemOpen
  \bibfield  {author} {\bibinfo {author} {\bibfnamefont {A.}~\bibnamefont {Banerjee}} \emph {et~al.},\ }\href {\doibase https://doi.org/10.1140/epjc/s10052-025-14596-x} {\bibfield  {journal} {\bibinfo  {journal} {Eur. Phys. J. C}\ }\textbf {\bibinfo {volume} {85}},\ \bibinfo {pages} {844} (\bibinfo {year} {2025})}\BibitemShut {NoStop}%
\bibitem [{\citenamefont {Panotopoulos}(2025)}]{panotopoulos2025UR}%
  \BibitemOpen
  \bibfield  {author} {\bibinfo {author} {\bibfnamefont {G.}~\bibnamefont {Panotopoulos}},\ }\href {\doibase https://doi.org/10.3390/galaxies13010013} {\bibfield  {journal} {\bibinfo  {journal} {Galaxies}\ }\textbf {\bibinfo {volume} {13}},\ \bibinfo {pages} {13} (\bibinfo {year} {2025})}\BibitemShut {NoStop}%
\bibitem [{\citenamefont {Dymnikova}(1992)}]{dymnikova1992vacuum}%
  \BibitemOpen
  \bibfield  {author} {\bibinfo {author} {\bibfnamefont {I.}~\bibnamefont {Dymnikova}},\ }\href {\doibase 10.1007/BF00760226} {\bibfield  {journal} {\bibinfo  {journal} {Gen. Relativ. Gravit.}\ }\textbf {\bibinfo {volume} {24}},\ \bibinfo {pages} {235} (\bibinfo {year} {1992})}\BibitemShut {NoStop}%
\bibitem [{\citenamefont {Coleman}\ and\ \citenamefont {De~Luccia}(1980)}]{coleman1980gravitational}%
  \BibitemOpen
  \bibfield  {author} {\bibinfo {author} {\bibfnamefont {S.}~\bibnamefont {Coleman}}\ and\ \bibinfo {author} {\bibfnamefont {F.}~\bibnamefont {De~Luccia}},\ }\href {\doibase 10.1103/PhysRevD.21.3305} {\bibfield  {journal} {\bibinfo  {journal} {Phys. Rev. D}\ }\textbf {\bibinfo {volume} {21}},\ \bibinfo {pages} {3305} (\bibinfo {year} {1980})}\BibitemShut {NoStop}%
\bibitem [{\citenamefont {Mazur}\ and\ \citenamefont {Mottola}(2004)}]{mazur2004gravitational}%
  \BibitemOpen
  \bibfield  {author} {\bibinfo {author} {\bibfnamefont {P.~O.}\ \bibnamefont {Mazur}}\ and\ \bibinfo {author} {\bibfnamefont {E.}~\bibnamefont {Mottola}},\ }\href {\doibase 10.1073/pnas.0402717101} {\bibfield  {journal} {\bibinfo  {journal} {Proceedings of the National Academy of Sciences}\ }\textbf {\bibinfo {volume} {101}},\ \bibinfo {pages} {9545} (\bibinfo {year} {2004})}\BibitemShut {NoStop}%
\bibitem [{\citenamefont {Mazur}\ and\ \citenamefont {Mottola}(2023)}]{mazur2023gravitational}%
  \BibitemOpen
  \bibfield  {author} {\bibinfo {author} {\bibfnamefont {P.~O.}\ \bibnamefont {Mazur}}\ and\ \bibinfo {author} {\bibfnamefont {E.}~\bibnamefont {Mottola}},\ }\href {\doibase 10.3390/universe9020088} {\bibfield  {journal} {\bibinfo  {journal} {Universe}\ }\textbf {\bibinfo {volume} {9}},\ \bibinfo {pages} {88} (\bibinfo {year} {2023})}\BibitemShut {NoStop}%
\bibitem [{\citenamefont {Kiselev}(2003)}]{Kiselev2003}%
  \BibitemOpen
  \bibfield  {author} {\bibinfo {author} {\bibfnamefont {V.~V.}\ \bibnamefont {Kiselev}},\ }\href {\doibase 10.1088/0264-9381/20/6/310} {\bibfield  {journal} {\bibinfo  {journal} {Class. Quantum Grav.}\ }\textbf {\bibinfo {volume} {20}},\ \bibinfo {pages} {1187} (\bibinfo {year} {2003})}\BibitemShut {NoStop}%
\bibitem [{\citenamefont {Fernando}(2012)}]{Fernando2012}%
  \BibitemOpen
  \bibfield  {author} {\bibinfo {author} {\bibfnamefont {S.}~\bibnamefont {Fernando}},\ }\href {\doibase 10.1007/s10714-012-1368-x} {\bibfield  {journal} {\bibinfo  {journal} {Gen. Relativ. Gravit.}\ }\textbf {\bibinfo {volume} {44}},\ \bibinfo {pages} {1857} (\bibinfo {year} {2012})}\BibitemShut {NoStop}%
\bibitem [{\citenamefont {Sushkov}(2005)}]{Sushkov2005Wormholes}%
  \BibitemOpen
  \bibfield  {author} {\bibinfo {author} {\bibfnamefont {S.~V.}\ \bibnamefont {Sushkov}},\ }\href {\doibase 10.1103/PhysRevD.71.043520} {\bibfield  {journal} {\bibinfo  {journal} {Phys. Rev. D}\ }\textbf {\bibinfo {volume} {71}},\ \bibinfo {pages} {043520} (\bibinfo {year} {2005})}\BibitemShut {NoStop}%
\bibitem [{\citenamefont {Chapline}(2005)}]{chapline2005dark}%
  \BibitemOpen
  \bibfield  {author} {\bibinfo {author} {\bibfnamefont {G.}~\bibnamefont {Chapline}},\ }\href {https://arxiv.org/abs/astro-ph/0503200} {\bibfield  {journal} {\bibinfo  {journal} {arXiv:astro-ph/0503200}\ } (\bibinfo {year} {2005})}\BibitemShut {NoStop}%
\bibitem [{\citenamefont {Beltracchi}\ and\ \citenamefont {Gondolo}(2019)}]{Beltracchi2019}%
  \BibitemOpen
  \bibfield  {author} {\bibinfo {author} {\bibfnamefont {P.}~\bibnamefont {Beltracchi}}\ and\ \bibinfo {author} {\bibfnamefont {P.}~\bibnamefont {Gondolo}},\ }\href {\doibase 10.1103/PhysRevD.99.044037} {\bibfield  {journal} {\bibinfo  {journal} {Phys. Rev. D}\ }\textbf {\bibinfo {volume} {99}},\ \bibinfo {pages} {044037} (\bibinfo {year} {2019})}\BibitemShut {NoStop}%
\bibitem [{\citenamefont {Yazadjiev}(2011)}]{yazadjiev2011exact}%
  \BibitemOpen
  \bibfield  {author} {\bibinfo {author} {\bibfnamefont {S.~S.}\ \bibnamefont {Yazadjiev}},\ }\href {\doibase 10.1103/PhysRevD.83.127501} {\bibfield  {journal} {\bibinfo  {journal} {Phys. Rev. D}\ }\textbf {\bibinfo {volume} {83}},\ \bibinfo {pages} {127501} (\bibinfo {year} {2011})}\BibitemShut {NoStop}%
\bibitem [{\citenamefont {Sakti}\ and\ \citenamefont {Sulaksono}(2021)}]{sakti2021dark}%
  \BibitemOpen
  \bibfield  {author} {\bibinfo {author} {\bibfnamefont {M.~F.}\ \bibnamefont {Sakti}}\ and\ \bibinfo {author} {\bibfnamefont {A.}~\bibnamefont {Sulaksono}},\ }\href {\doibase 10.1103/PhysRevD.103.084042} {\bibfield  {journal} {\bibinfo  {journal} {Phys. Rev. D}\ }\textbf {\bibinfo {volume} {103}},\ \bibinfo {pages} {084042} (\bibinfo {year} {2021})}\BibitemShut {NoStop}%
\bibitem [{\citenamefont {Ghezzi}(2011)}]{ghezzi2011anisotropic}%
  \BibitemOpen
  \bibfield  {author} {\bibinfo {author} {\bibfnamefont {C.~R.}\ \bibnamefont {Ghezzi}},\ }\href {\doibase 10.1007/s10509-011-0663-4} {\bibfield  {journal} {\bibinfo  {journal} {Astrophys. Space Sci.}\ }\textbf {\bibinfo {volume} {333}},\ \bibinfo {pages} {437} (\bibinfo {year} {2011})}\BibitemShut {NoStop}%
\bibitem [{\citenamefont {Tudeshki}\ \emph {et~al.}(2022)\citenamefont {Tudeshki}, \citenamefont {Bordbar},\ and\ \citenamefont {Panah}}]{tudeshki2022dark}%
  \BibitemOpen
  \bibfield  {author} {\bibinfo {author} {\bibfnamefont {A.~B.}\ \bibnamefont {Tudeshki}}, \bibinfo {author} {\bibfnamefont {G.}~\bibnamefont {Bordbar}}, \ and\ \bibinfo {author} {\bibfnamefont {B.~E.}\ \bibnamefont {Panah}},\ }\href {\doibase 10.1016/j.physletb.2022.137523} {\bibfield  {journal} {\bibinfo  {journal} {Phys. Lett. B}\ }\textbf {\bibinfo {volume} {835}},\ \bibinfo {pages} {137523} (\bibinfo {year} {2022})}\BibitemShut {NoStop}%
\bibitem [{\citenamefont {Tudeshki}\ \emph {et~al.}(2024)\citenamefont {Tudeshki}, \citenamefont {Bordbar},\ and\ \citenamefont {Panah}}]{tudeshki2024effect}%
  \BibitemOpen
  \bibfield  {author} {\bibinfo {author} {\bibfnamefont {A.~B.}\ \bibnamefont {Tudeshki}}, \bibinfo {author} {\bibfnamefont {G.}~\bibnamefont {Bordbar}}, \ and\ \bibinfo {author} {\bibfnamefont {B.~E.}\ \bibnamefont {Panah}},\ }\href {\doibase 10.1016/j.physletb.2023.138333} {\bibfield  {journal} {\bibinfo  {journal} {Phys. Lett. B}\ }\textbf {\bibinfo {volume} {848}},\ \bibinfo {pages} {138333} (\bibinfo {year} {2024})}\BibitemShut {NoStop}%
\bibitem [{\citenamefont {Das}\ and\ \citenamefont {Debnath}(2024)}]{das2024effect}%
  \BibitemOpen
  \bibfield  {author} {\bibinfo {author} {\bibfnamefont {K.~P.}\ \bibnamefont {Das}}\ and\ \bibinfo {author} {\bibfnamefont {U.}~\bibnamefont {Debnath}},\ }\href {\doibase 10.1140/epjp/s13360-024-05785-w} {\bibfield  {journal} {\bibinfo  {journal} {Eur. Phys. J. Plus}\ }\textbf {\bibinfo {volume} {139}},\ \bibinfo {pages} {988} (\bibinfo {year} {2024})}\BibitemShut {NoStop}%
\bibitem [{\citenamefont {Smerechynskyi}\ \emph {et~al.}(2021)\citenamefont {Smerechynskyi}, \citenamefont {Tsizh},\ and\ \citenamefont {Novosyadlyj}}]{smerechynskyi2021impact}%
  \BibitemOpen
  \bibfield  {author} {\bibinfo {author} {\bibfnamefont {S.}~\bibnamefont {Smerechynskyi}}, \bibinfo {author} {\bibfnamefont {M.}~\bibnamefont {Tsizh}}, \ and\ \bibinfo {author} {\bibfnamefont {B.}~\bibnamefont {Novosyadlyj}},\ }\href {\doibase 10.1088/1475-7516/2021/02/045} {\bibfield  {journal} {\bibinfo  {journal} {JCAP}\ }\textbf {\bibinfo {volume} {02}},\ \bibinfo {pages} {045} (\bibinfo {year} {2021})}\BibitemShut {NoStop}%
\bibitem [{\citenamefont {Pretel}\ \emph {et~al.}(2024{\natexlab{a}})\citenamefont {Pretel}, \citenamefont {Dutra},\ and\ \citenamefont {Duarte}}]{Pretel2024PRD1}%
  \BibitemOpen
  \bibfield  {author} {\bibinfo {author} {\bibfnamefont {J.~M.~Z.}\ \bibnamefont {Pretel}}, \bibinfo {author} {\bibfnamefont {M.}~\bibnamefont {Dutra}}, \ and\ \bibinfo {author} {\bibfnamefont {S.~B.}\ \bibnamefont {Duarte}},\ }\href {\doibase 10.1103/PhysRevD.109.023524} {\bibfield  {journal} {\bibinfo  {journal} {Phys. Rev. D}\ }\textbf {\bibinfo {volume} {109}},\ \bibinfo {pages} {023524} (\bibinfo {year} {2024}{\natexlab{a}})}\BibitemShut {NoStop}%
\bibitem [{\citenamefont {Pretel}\ \emph {et~al.}(2024{\natexlab{b}})\citenamefont {Pretel}, \citenamefont {Duarte}, \citenamefont {Arba\~nil}, \citenamefont {Dutra},\ and\ \citenamefont {Louren\ifmmode~\mbox{\c{c}}\else \c{c}\fi{}o}}]{Pretel2024PRD2}%
  \BibitemOpen
  \bibfield  {author} {\bibinfo {author} {\bibfnamefont {J.~M.~Z.}\ \bibnamefont {Pretel}}, \bibinfo {author} {\bibfnamefont {S.~B.}\ \bibnamefont {Duarte}}, \bibinfo {author} {\bibfnamefont {J.~D.~V.}\ \bibnamefont {Arba\~nil}}, \bibinfo {author} {\bibfnamefont {M.}~\bibnamefont {Dutra}}, \ and\ \bibinfo {author} {\bibfnamefont {O.}~\bibnamefont {Louren\ifmmode~\mbox{\c{c}}\else \c{c}\fi{}o}},\ }\href {\doibase 10.1103/PhysRevD.110.124019} {\bibfield  {journal} {\bibinfo  {journal} {Phys. Rev. D}\ }\textbf {\bibinfo {volume} {110}},\ \bibinfo {pages} {124019} (\bibinfo {year} {2024}{\natexlab{b}})}\BibitemShut {NoStop}%
\bibitem [{\citenamefont {{Pretel}}\ \emph {et~al.}(2024)\citenamefont {{Pretel}}, \citenamefont {{Dutra}},\ and\ \citenamefont {{Duarte}}}]{2024arXiv241213568P}%
  \BibitemOpen
  \bibfield  {author} {\bibinfo {author} {\bibfnamefont {J.~M.~Z.}\ \bibnamefont {{Pretel}}}, \bibinfo {author} {\bibfnamefont {M.}~\bibnamefont {{Dutra}}}, \ and\ \bibinfo {author} {\bibfnamefont {S.~B.}\ \bibnamefont {{Duarte}}},\ }\href@noop {} {\  (\bibinfo {year} {2024})},\ \Eprint {http://arxiv.org/abs/2412.13568} {arXiv:2412.13568 [gr-qc]} \BibitemShut {NoStop}%
\bibitem [{\citenamefont {Chan}\ \emph {et~al.}(2009)\citenamefont {Chan}, \citenamefont {Da~Silva},\ and\ \citenamefont {Villas~da Rocha}}]{chan2009star}%
  \BibitemOpen
  \bibfield  {author} {\bibinfo {author} {\bibfnamefont {R.}~\bibnamefont {Chan}}, \bibinfo {author} {\bibfnamefont {M.}~\bibnamefont {Da~Silva}}, \ and\ \bibinfo {author} {\bibfnamefont {J.~F.}\ \bibnamefont {Villas~da Rocha}},\ }\href {\doibase 10.1007/s10714-008-0755-9} {\bibfield  {journal} {\bibinfo  {journal} {Gen. Relativ. Gravit.}\ }\textbf {\bibinfo {volume} {41}},\ \bibinfo {pages} {1835} (\bibinfo {year} {2009})}\BibitemShut {NoStop}%
\bibitem [{\citenamefont {Bhar}\ \emph {et~al.}(2018)\citenamefont {Bhar}, \citenamefont {Govender},\ and\ \citenamefont {Sharma}}]{bhar2018anisotropic}%
  \BibitemOpen
  \bibfield  {author} {\bibinfo {author} {\bibfnamefont {P.}~\bibnamefont {Bhar}}, \bibinfo {author} {\bibfnamefont {M.}~\bibnamefont {Govender}}, \ and\ \bibinfo {author} {\bibfnamefont {R.}~\bibnamefont {Sharma}},\ }\href {\doibase 10.1007/s12043-017-1500-2} {\bibfield  {journal} {\bibinfo  {journal} {Pramana}\ }\textbf {\bibinfo {volume} {90}},\ \bibinfo {pages} {1} (\bibinfo {year} {2018})}\BibitemShut {NoStop}%
\bibitem [{\citenamefont {Estevez-Delgado}\ \emph {et~al.}(2021)\citenamefont {Estevez-Delgado}, \citenamefont {Duran}, \citenamefont {Cleary-Balderas}, \citenamefont {Rodr{\'\i}guez~Maya},\ and\ \citenamefont {Pe{\~n}a}}]{estevez2021chaplygin}%
  \BibitemOpen
  \bibfield  {author} {\bibinfo {author} {\bibfnamefont {J.}~\bibnamefont {Estevez-Delgado}}, \bibinfo {author} {\bibfnamefont {M.~P.}\ \bibnamefont {Duran}}, \bibinfo {author} {\bibfnamefont {A.}~\bibnamefont {Cleary-Balderas}}, \bibinfo {author} {\bibfnamefont {N.~E.}\ \bibnamefont {Rodr{\'\i}guez~Maya}}, \ and\ \bibinfo {author} {\bibfnamefont {J.~M.}\ \bibnamefont {Pe{\~n}a}},\ }\href {\doibase 10.1142/S0217732321502138} {\bibfield  {journal} {\bibinfo  {journal} {Mod. Phys. Lett. A}\ }\textbf {\bibinfo {volume} {36}},\ \bibinfo {pages} {2150213} (\bibinfo {year} {2021})}\BibitemShut {NoStop}%
\bibitem [{\citenamefont {Das}\ \emph {et~al.}(2023)\citenamefont {Das}, \citenamefont {Debnath},\ and\ \citenamefont {Ray}}]{das2023dark}%
  \BibitemOpen
  \bibfield  {author} {\bibinfo {author} {\bibfnamefont {K.~P.}\ \bibnamefont {Das}}, \bibinfo {author} {\bibfnamefont {U.}~\bibnamefont {Debnath}}, \ and\ \bibinfo {author} {\bibfnamefont {S.}~\bibnamefont {Ray}},\ }\href {\doibase 10.1002/prop.202200148} {\bibfield  {journal} {\bibinfo  {journal} {Fortschr. Phys.}\ }\textbf {\bibinfo {volume} {71}},\ \bibinfo {pages} {2200148} (\bibinfo {year} {2023})}\BibitemShut {NoStop}%
\bibitem [{\citenamefont {Bhar}\ and\ \citenamefont {Pretel}(2023)}]{bhar2023dark}%
  \BibitemOpen
  \bibfield  {author} {\bibinfo {author} {\bibfnamefont {P.}~\bibnamefont {Bhar}}\ and\ \bibinfo {author} {\bibfnamefont {J.~M.~Z.}\ \bibnamefont {Pretel}},\ }\href {\doibase 10.1016/j.dark.2023.101322} {\bibfield  {journal} {\bibinfo  {journal} {Phys. Dark Univ.}\ }\textbf {\bibinfo {volume} {42}},\ \bibinfo {pages} {101322} (\bibinfo {year} {2023})}\BibitemShut {NoStop}%
\bibitem [{\citenamefont {Das}\ \emph {et~al.}(2024)\citenamefont {Das}, \citenamefont {Debnath}, \citenamefont {Ashraf},\ and\ \citenamefont {Khurana}}]{das2024acceptable}%
  \BibitemOpen
  \bibfield  {author} {\bibinfo {author} {\bibfnamefont {K.~P.}\ \bibnamefont {Das}}, \bibinfo {author} {\bibfnamefont {U.}~\bibnamefont {Debnath}}, \bibinfo {author} {\bibfnamefont {A.}~\bibnamefont {Ashraf}}, \ and\ \bibinfo {author} {\bibfnamefont {M.}~\bibnamefont {Khurana}},\ }\href {\doibase 10.1016/j.dark.2023.101398} {\bibfield  {journal} {\bibinfo  {journal} {Phys. Dark Univ.}\ }\textbf {\bibinfo {volume} {43}},\ \bibinfo {pages} {101398} (\bibinfo {year} {2024})}\BibitemShut {NoStop}%
\bibitem [{\citenamefont {Das}\ and\ \citenamefont {Debnath}(2025)}]{das2025study}%
  \BibitemOpen
  \bibfield  {author} {\bibinfo {author} {\bibfnamefont {K.~P.}\ \bibnamefont {Das}}\ and\ \bibinfo {author} {\bibfnamefont {U.}~\bibnamefont {Debnath}},\ }\href {\doibase 10.1140/epjc/s10052-025-14059-3} {\bibfield  {journal} {\bibinfo  {journal} {Eur. Phys. J. C}\ }\textbf {\bibinfo {volume} {85}},\ \bibinfo {pages} {329} (\bibinfo {year} {2025})}\BibitemShut {NoStop}%
\bibitem [{\citenamefont {D'Onofrio}(2025)}]{Onofrio2025}%
  \BibitemOpen
  \bibfield  {author} {\bibinfo {author} {\bibfnamefont {S.}~\bibnamefont {D'Onofrio}},\ }\href {\doibase 10.1140/epjc/s10052-025-14397-2} {\bibfield  {journal} {\bibinfo  {journal} {Eur. Phys. J. C}\ }\textbf {\bibinfo {volume} {85}},\ \bibinfo {pages} {672} (\bibinfo {year} {2025})}\BibitemShut {NoStop}%
\bibitem [{\citenamefont {Yagi}\ and\ \citenamefont {Yunes}(2013{\natexlab{a}})}]{yagi2013love}%
  \BibitemOpen
  \bibfield  {author} {\bibinfo {author} {\bibfnamefont {K.}~\bibnamefont {Yagi}}\ and\ \bibinfo {author} {\bibfnamefont {N.}~\bibnamefont {Yunes}},\ }\href {\doibase 10.1103/PhysRevD.88.023009} {\bibfield  {journal} {\bibinfo  {journal} {Phys. Rev. D}\ }\textbf {\bibinfo {volume} {88}},\ \bibinfo {pages} {023009} (\bibinfo {year} {2013}{\natexlab{a}})}\BibitemShut {NoStop}%
\bibitem [{\citenamefont {Yagi}\ and\ \citenamefont {Yunes}(2015)}]{yagi2015love}%
  \BibitemOpen
  \bibfield  {author} {\bibinfo {author} {\bibfnamefont {K.}~\bibnamefont {Yagi}}\ and\ \bibinfo {author} {\bibfnamefont {N.}~\bibnamefont {Yunes}},\ }\href {\doibase 10.1103/PhysRevD.91.123008} {\bibfield  {journal} {\bibinfo  {journal} {Phys. Rev. D}\ }\textbf {\bibinfo {volume} {91}},\ \bibinfo {pages} {123008} (\bibinfo {year} {2015})}\BibitemShut {NoStop}%
\bibitem [{\citenamefont {Haskell}\ \emph {et~al.}(2013)\citenamefont {Haskell}, \citenamefont {Ciolfi}, \citenamefont {Pannarale},\ and\ \citenamefont {Rezzolla}}]{haskell2013universality}%
  \BibitemOpen
  \bibfield  {author} {\bibinfo {author} {\bibfnamefont {B.}~\bibnamefont {Haskell}}, \bibinfo {author} {\bibfnamefont {R.}~\bibnamefont {Ciolfi}}, \bibinfo {author} {\bibfnamefont {F.}~\bibnamefont {Pannarale}}, \ and\ \bibinfo {author} {\bibfnamefont {L.}~\bibnamefont {Rezzolla}},\ }\href {\doibase 10.1093/mnrasl/slt161} {\bibfield  {journal} {\bibinfo  {journal} {MNRAS}\ }\textbf {\bibinfo {volume} {438}},\ \bibinfo {pages} {L71} (\bibinfo {year} {2013})}\BibitemShut {NoStop}%
\bibitem [{\citenamefont {Chakrabarti}\ \emph {et~al.}(2014)\citenamefont {Chakrabarti}, \citenamefont {Delsate}, \citenamefont {G{\"u}rlebeck},\ and\ \citenamefont {Steinhoff}}]{chakrabarti2014q}%
  \BibitemOpen
  \bibfield  {author} {\bibinfo {author} {\bibfnamefont {S.}~\bibnamefont {Chakrabarti}}, \bibinfo {author} {\bibfnamefont {T.}~\bibnamefont {Delsate}}, \bibinfo {author} {\bibfnamefont {N.}~\bibnamefont {G{\"u}rlebeck}}, \ and\ \bibinfo {author} {\bibfnamefont {J.}~\bibnamefont {Steinhoff}},\ }\href {\doibase 10.1103/PhysRevLett.112.201102} {\bibfield  {journal} {\bibinfo  {journal} {Phys. Rev. Lett.}\ }\textbf {\bibinfo {volume} {112}},\ \bibinfo {pages} {201102} (\bibinfo {year} {2014})}\BibitemShut {NoStop}%
\bibitem [{\citenamefont {Gupta}\ \emph {et~al.}(2017)\citenamefont {Gupta}, \citenamefont {Majumder}, \citenamefont {Yagi},\ and\ \citenamefont {Yunes}}]{gupta2017love}%
  \BibitemOpen
  \bibfield  {author} {\bibinfo {author} {\bibfnamefont {T.}~\bibnamefont {Gupta}}, \bibinfo {author} {\bibfnamefont {B.}~\bibnamefont {Majumder}}, \bibinfo {author} {\bibfnamefont {K.}~\bibnamefont {Yagi}}, \ and\ \bibinfo {author} {\bibfnamefont {N.}~\bibnamefont {Yunes}},\ }\href {\doibase 10.1088/1361-6382/aa9c68} {\bibfield  {journal} {\bibinfo  {journal} {Class. Quantum Grav.}\ }\textbf {\bibinfo {volume} {35}},\ \bibinfo {pages} {025009} (\bibinfo {year} {2017})}\BibitemShut {NoStop}%
\bibitem [{\citenamefont {Bandyopadhyay}\ \emph {et~al.}(2018)\citenamefont {Bandyopadhyay}, \citenamefont {Bhat}, \citenamefont {Char},\ and\ \citenamefont {Chatterjee}}]{bandyopadhyay2018moment}%
  \BibitemOpen
  \bibfield  {author} {\bibinfo {author} {\bibfnamefont {D.}~\bibnamefont {Bandyopadhyay}}, \bibinfo {author} {\bibfnamefont {S.~A.}\ \bibnamefont {Bhat}}, \bibinfo {author} {\bibfnamefont {P.}~\bibnamefont {Char}}, \ and\ \bibinfo {author} {\bibfnamefont {D.}~\bibnamefont {Chatterjee}},\ }\href {\doibase 10.1140/epja/i2018-12456-y} {\bibfield  {journal} {\bibinfo  {journal} {Eur. Phys. J. A}\ }\textbf {\bibinfo {volume} {54}},\ \bibinfo {pages} {26} (\bibinfo {year} {2018})}\BibitemShut {NoStop}%
\bibitem [{\citenamefont {Jiang}\ \emph {et~al.}(2020)\citenamefont {Jiang}, \citenamefont {Tang}, \citenamefont {Wang}, \citenamefont {Fan},\ and\ \citenamefont {Wei}}]{jiang2020psr}%
  \BibitemOpen
  \bibfield  {author} {\bibinfo {author} {\bibfnamefont {J.-L.}\ \bibnamefont {Jiang}}, \bibinfo {author} {\bibfnamefont {S.-P.}\ \bibnamefont {Tang}}, \bibinfo {author} {\bibfnamefont {Y.-Z.}\ \bibnamefont {Wang}}, \bibinfo {author} {\bibfnamefont {Y.-Z.}\ \bibnamefont {Fan}}, \ and\ \bibinfo {author} {\bibfnamefont {D.-M.}\ \bibnamefont {Wei}},\ }\href {\doibase 10.3847/1538-4357/ab77cf} {\bibfield  {journal} {\bibinfo  {journal} {Astrophys. J.}\ }\textbf {\bibinfo {volume} {892}},\ \bibinfo {pages} {55} (\bibinfo {year} {2020})}\BibitemShut {NoStop}%
\bibitem [{\citenamefont {Yeung}\ \emph {et~al.}(2021)\citenamefont {Yeung}, \citenamefont {Lin}, \citenamefont {Andersson},\ and\ \citenamefont {Comer}}]{yeung2021love}%
  \BibitemOpen
  \bibfield  {author} {\bibinfo {author} {\bibfnamefont {C.-H.}\ \bibnamefont {Yeung}}, \bibinfo {author} {\bibfnamefont {L.-M.}\ \bibnamefont {Lin}}, \bibinfo {author} {\bibfnamefont {N.}~\bibnamefont {Andersson}}, \ and\ \bibinfo {author} {\bibfnamefont {G.}~\bibnamefont {Comer}},\ }\href {\doibase 10.3390/universe7040111} {\bibfield  {journal} {\bibinfo  {journal} {Universe}\ }\textbf {\bibinfo {volume} {7}},\ \bibinfo {pages} {111} (\bibinfo {year} {2021})}\BibitemShut {NoStop}%
\bibitem [{\citenamefont {Zhao}\ and\ \citenamefont {Lattimer}(2022)}]{Zhao2022}%
  \BibitemOpen
  \bibfield  {author} {\bibinfo {author} {\bibfnamefont {T.}~\bibnamefont {Zhao}}\ and\ \bibinfo {author} {\bibfnamefont {J.~M.}\ \bibnamefont {Lattimer}},\ }\href {\doibase 10.1103/PhysRevD.106.123002} {\bibfield  {journal} {\bibinfo  {journal} {Phys. Rev. D}\ }\textbf {\bibinfo {volume} {106}},\ \bibinfo {pages} {123002} (\bibinfo {year} {2022})}\BibitemShut {NoStop}%
\bibitem [{\citenamefont {Pretel}(2024)}]{Pretel2024PS}%
  \BibitemOpen
  \bibfield  {author} {\bibinfo {author} {\bibfnamefont {J.~M.~Z.}\ \bibnamefont {Pretel}},\ }\href {\doibase 10.1088/1402-4896/ad5ac4} {\bibfield  {journal} {\bibinfo  {journal} {Phys. Scr.}\ }\textbf {\bibinfo {volume} {99}},\ \bibinfo {pages} {085001} (\bibinfo {year} {2024})}\BibitemShut {NoStop}%
\bibitem [{\citenamefont {Negreiros}\ \emph {et~al.}(2025)\citenamefont {Negreiros}, \citenamefont {Zhang},\ and\ \citenamefont {Xu}}]{Negreiros2025}%
  \BibitemOpen
  \bibfield  {author} {\bibinfo {author} {\bibfnamefont {R.}~\bibnamefont {Negreiros}}, \bibinfo {author} {\bibfnamefont {C.}~\bibnamefont {Zhang}}, \ and\ \bibinfo {author} {\bibfnamefont {R.}~\bibnamefont {Xu}},\ }\href {\doibase 10.1103/PhysRevD.111.063026} {\bibfield  {journal} {\bibinfo  {journal} {Phys. Rev. D}\ }\textbf {\bibinfo {volume} {111}},\ \bibinfo {pages} {063026} (\bibinfo {year} {2025})}\BibitemShut {NoStop}%
\bibitem [{\citenamefont {Das}(2022)}]{das2022love}%
  \BibitemOpen
  \bibfield  {author} {\bibinfo {author} {\bibfnamefont {H.~C.}\ \bibnamefont {Das}},\ }\href {\doibase 10.1103/PhysRevD.106.103518} {\bibfield  {journal} {\bibinfo  {journal} {Phys. Rev. D}\ }\textbf {\bibinfo {volume} {106}},\ \bibinfo {pages} {103518} (\bibinfo {year} {2022})}\BibitemShut {NoStop}%
\bibitem [{\citenamefont {Mohanty}\ \emph {et~al.}(2024)\citenamefont {Mohanty}, \citenamefont {Ghosh}, \citenamefont {Routaray}, \citenamefont {Das},\ and\ \citenamefont {Kumar}}]{mohanty2024impact}%
  \BibitemOpen
  \bibfield  {author} {\bibinfo {author} {\bibfnamefont {S.~R.}\ \bibnamefont {Mohanty}}, \bibinfo {author} {\bibfnamefont {S.}~\bibnamefont {Ghosh}}, \bibinfo {author} {\bibfnamefont {P.}~\bibnamefont {Routaray}}, \bibinfo {author} {\bibfnamefont {H.}~\bibnamefont {Das}}, \ and\ \bibinfo {author} {\bibfnamefont {B.}~\bibnamefont {Kumar}},\ }\href {\doibase 10.1088/1475-7516/2024/03/054} {\bibfield  {journal} {\bibinfo  {journal} {JCAP}\ }\textbf {\bibinfo {volume} {03}},\ \bibinfo {pages} {054} (\bibinfo {year} {2024})}\BibitemShut {NoStop}%
\bibitem [{\citenamefont {Guedes}\ \emph {et~al.}(2026)\citenamefont {Guedes}, \citenamefont {Ajith}, \citenamefont {Lau},\ and\ \citenamefont {Yagi}}]{Guedes2026}%
  \BibitemOpen
  \bibfield  {author} {\bibinfo {author} {\bibfnamefont {V.}~\bibnamefont {Guedes}}, \bibinfo {author} {\bibfnamefont {S.}~\bibnamefont {Ajith}}, \bibinfo {author} {\bibfnamefont {S.~Y.}\ \bibnamefont {Lau}}, \ and\ \bibinfo {author} {\bibfnamefont {K.}~\bibnamefont {Yagi}},\ }\href {\doibase 10.1103/s9zp-jfnh} {\bibfield  {journal} {\bibinfo  {journal} {Phys. Rev. D}\ }\textbf {\bibinfo {volume} {113}},\ \bibinfo {pages} {043025} (\bibinfo {year} {2026})}\BibitemShut {NoStop}%
\bibitem [{\citenamefont {Sotani}\ and\ \citenamefont {Kumar}(2025)}]{Sotani2025}%
  \BibitemOpen
  \bibfield  {author} {\bibinfo {author} {\bibfnamefont {H.}~\bibnamefont {Sotani}}\ and\ \bibinfo {author} {\bibfnamefont {A.}~\bibnamefont {Kumar}},\ }\href {\doibase 10.1140/epjc/s10052-025-15186-7} {\bibfield  {journal} {\bibinfo  {journal} {Eur. Phys. J. C}\ }\textbf {\bibinfo {volume} {85}},\ \bibinfo {pages} {1438} (\bibinfo {year} {2025})}\BibitemShut {NoStop}%
\bibitem [{\citenamefont {Doroshenko}\ \emph {et~al.}(2022)\citenamefont {Doroshenko}, \citenamefont {Suleimanov}, \citenamefont {Phlhofer},\ and\ \citenamefont {Santangelo}}]{Doroshenko2022}%
  \BibitemOpen
  \bibfield  {author} {\bibinfo {author} {\bibfnamefont {V.}~\bibnamefont {Doroshenko}}, \bibinfo {author} {\bibfnamefont {V.}~\bibnamefont {Suleimanov}}, \bibinfo {author} {\bibfnamefont {G.}~\bibnamefont {Phlhofer}}, \ and\ \bibinfo {author} {\bibfnamefont {A.}~\bibnamefont {Santangelo}},\ }\href {\doibase 10.1038/s41550-022-01800-1} {\bibfield  {journal} {\bibinfo  {journal} {Nat. Astron.}\ }\textbf {\bibinfo {volume} {6}},\ \bibinfo {pages} {1444} (\bibinfo {year} {2022})}\BibitemShut {NoStop}%
\bibitem [{\citenamefont {Riley}\ \emph {et~al.}(2019)\citenamefont {Riley} \emph {et~al.}}]{Riley2019}%
  \BibitemOpen
  \bibfield  {author} {\bibinfo {author} {\bibfnamefont {T.~E.}\ \bibnamefont {Riley}} \emph {et~al.},\ }\href {\doibase 10.3847/2041-8213/ab481c} {\bibfield  {journal} {\bibinfo  {journal} {Astrophys. J. Lett.}\ }\textbf {\bibinfo {volume} {887}},\ \bibinfo {pages} {L21} (\bibinfo {year} {2019})}\BibitemShut {NoStop}%
\bibitem [{\citenamefont {Miller}\ \emph {et~al.}(2019)\citenamefont {Miller} \emph {et~al.}}]{Miller2019}%
  \BibitemOpen
  \bibfield  {author} {\bibinfo {author} {\bibfnamefont {M.~C.}\ \bibnamefont {Miller}} \emph {et~al.},\ }\href {\doibase 10.3847/2041-8213/ab50c5} {\bibfield  {journal} {\bibinfo  {journal} {Astrophys. J. Lett.}\ }\textbf {\bibinfo {volume} {887}},\ \bibinfo {pages} {L24} (\bibinfo {year} {2019})}\BibitemShut {NoStop}%
\bibitem [{\citenamefont {Riley}\ \emph {et~al.}(2021)\citenamefont {Riley} \emph {et~al.}}]{Riley2021}%
  \BibitemOpen
  \bibfield  {author} {\bibinfo {author} {\bibfnamefont {T.~E.}\ \bibnamefont {Riley}} \emph {et~al.},\ }\href {\doibase 10.3847/2041-8213/ac0a81} {\bibfield  {journal} {\bibinfo  {journal} {Astrophys. J. Lett.}\ }\textbf {\bibinfo {volume} {918}},\ \bibinfo {pages} {L27} (\bibinfo {year} {2021})}\BibitemShut {NoStop}%
\bibitem [{\citenamefont {Miller}\ \emph {et~al.}(2021)\citenamefont {Miller} \emph {et~al.}}]{Miller2021}%
  \BibitemOpen
  \bibfield  {author} {\bibinfo {author} {\bibfnamefont {M.~C.}\ \bibnamefont {Miller}} \emph {et~al.},\ }\href {\doibase 10.3847/2041-8213/ac089b} {\bibfield  {journal} {\bibinfo  {journal} {Astrophys. J. Lett.}\ }\textbf {\bibinfo {volume} {918}},\ \bibinfo {pages} {L28} (\bibinfo {year} {2021})}\BibitemShut {NoStop}%
\bibitem [{\citenamefont {Oppenheimer}\ and\ \citenamefont {Volkoff}(1939)}]{Oppenheimer:1939ne}%
  \BibitemOpen
  \bibfield  {author} {\bibinfo {author} {\bibfnamefont {J.~R.}\ \bibnamefont {Oppenheimer}}\ and\ \bibinfo {author} {\bibfnamefont {G.~M.}\ \bibnamefont {Volkoff}},\ }\href {\doibase 10.1103/PhysRev.55.374} {\bibfield  {journal} {\bibinfo  {journal} {Phys. Rev.}\ }\textbf {\bibinfo {volume} {55}},\ \bibinfo {pages} {374} (\bibinfo {year} {1939})}\BibitemShut {NoStop}%
\bibitem [{\citenamefont {Tolman}(1939)}]{Tolman:1939jz}%
  \BibitemOpen
  \bibfield  {author} {\bibinfo {author} {\bibfnamefont {R.~C.}\ \bibnamefont {Tolman}},\ }\href {\doibase 10.1103/PhysRev.55.364} {\bibfield  {journal} {\bibinfo  {journal} {Phys. Rev.}\ }\textbf {\bibinfo {volume} {55}},\ \bibinfo {pages} {364} (\bibinfo {year} {1939})}\BibitemShut {NoStop}%
\bibitem [{\citenamefont {Silva}\ \emph {et~al.}(2021)\citenamefont {Silva}, \citenamefont {Holgado}, \citenamefont {C\'ardenas-Avenda\~no},\ and\ \citenamefont {Yunes}}]{Silva2021}%
  \BibitemOpen
  \bibfield  {author} {\bibinfo {author} {\bibfnamefont {H.~O.}\ \bibnamefont {Silva}}, \bibinfo {author} {\bibfnamefont {A.~M.}\ \bibnamefont {Holgado}}, \bibinfo {author} {\bibfnamefont {A.}~\bibnamefont {C\'ardenas-Avenda\~no}}, \ and\ \bibinfo {author} {\bibfnamefont {N.}~\bibnamefont {Yunes}},\ }\href {\doibase 10.1103/PhysRevLett.126.181101} {\bibfield  {journal} {\bibinfo  {journal} {Phys. Rev. Lett.}\ }\textbf {\bibinfo {volume} {126}},\ \bibinfo {pages} {181101} (\bibinfo {year} {2021})}\BibitemShut {NoStop}%
\bibitem [{\citenamefont {Landry}\ and\ \citenamefont {Kumar}(2018)}]{Landry2018}%
  \BibitemOpen
  \bibfield  {author} {\bibinfo {author} {\bibfnamefont {P.}~\bibnamefont {Landry}}\ and\ \bibinfo {author} {\bibfnamefont {B.}~\bibnamefont {Kumar}},\ }\href {\doibase 10.3847/2041-8213/aaee76} {\bibfield  {journal} {\bibinfo  {journal} {Astrophys. J. Lett.}\ }\textbf {\bibinfo {volume} {868}},\ \bibinfo {pages} {L22} (\bibinfo {year} {2018})}\BibitemShut {NoStop}%
\bibitem [{\citenamefont {Hu}\ and\ \citenamefont {Freire}(2024)}]{Hu2024}%
  \BibitemOpen
  \bibfield  {author} {\bibinfo {author} {\bibfnamefont {H.}~\bibnamefont {Hu}}\ and\ \bibinfo {author} {\bibfnamefont {P.~C.~C.}\ \bibnamefont {Freire}},\ }\href {\doibase 10.3390/universe10040160} {\bibfield  {journal} {\bibinfo  {journal} {Universe}\ }\textbf {\bibinfo {volume} {10}},\ \bibinfo {pages} {160} (\bibinfo {year} {2024})}\BibitemShut {NoStop}%
\bibitem [{\citenamefont {Hartle}(1967)}]{hartle1967slowly}%
  \BibitemOpen
  \bibfield  {author} {\bibinfo {author} {\bibfnamefont {J.~B.}\ \bibnamefont {Hartle}},\ }\href {\doibase 10.1086/149400} {\bibfield  {journal} {\bibinfo  {journal} {Astrophys. J.}\ }\textbf {\bibinfo {volume} {150}},\ \bibinfo {pages} {1005} (\bibinfo {year} {1967})}\BibitemShut {NoStop}%
\bibitem [{\citenamefont {Thorne}\ and\ \citenamefont {Campolattaro}(1967)}]{thorne1967non}%
  \BibitemOpen
  \bibfield  {author} {\bibinfo {author} {\bibfnamefont {K.~S.}\ \bibnamefont {Thorne}}\ and\ \bibinfo {author} {\bibfnamefont {A.}~\bibnamefont {Campolattaro}},\ }\href {\doibase 10.1086/149288} {\bibfield  {journal} {\bibinfo  {journal} {Astrophys. J.}\ }\textbf {\bibinfo {volume} {149}},\ \bibinfo {pages} {591} (\bibinfo {year} {1967})}\BibitemShut {NoStop}%
\bibitem [{\citenamefont {Hinderer}(2008)}]{hinderer2008tidal}%
  \BibitemOpen
  \bibfield  {author} {\bibinfo {author} {\bibfnamefont {T.}~\bibnamefont {Hinderer}},\ }\href {\doibase 10.1086/533487} {\bibfield  {journal} {\bibinfo  {journal} {Astrophys. J.}\ }\textbf {\bibinfo {volume} {677}},\ \bibinfo {pages} {1216} (\bibinfo {year} {2008})}\BibitemShut {NoStop}%
\bibitem [{\citenamefont {Hinderer}\ \emph {et~al.}(2010)\citenamefont {Hinderer}, \citenamefont {Lackey}, \citenamefont {Lang},\ and\ \citenamefont {Read}}]{Hinderer2010}%
  \BibitemOpen
  \bibfield  {author} {\bibinfo {author} {\bibfnamefont {T.}~\bibnamefont {Hinderer}}, \bibinfo {author} {\bibfnamefont {B.~D.}\ \bibnamefont {Lackey}}, \bibinfo {author} {\bibfnamefont {R.~N.}\ \bibnamefont {Lang}}, \ and\ \bibinfo {author} {\bibfnamefont {J.~S.}\ \bibnamefont {Read}},\ }\href {\doibase 10.1103/PhysRevD.81.123016} {\bibfield  {journal} {\bibinfo  {journal} {Phys. Rev. D}\ }\textbf {\bibinfo {volume} {81}},\ \bibinfo {pages} {123016} (\bibinfo {year} {2010})}\BibitemShut {NoStop}%
\bibitem [{\citenamefont {Pretel}\ and\ \citenamefont {Zhang}(2024)}]{Pretel2024}%
  \BibitemOpen
  \bibfield  {author} {\bibinfo {author} {\bibfnamefont {J.~M.~Z.}\ \bibnamefont {Pretel}}\ and\ \bibinfo {author} {\bibfnamefont {C.}~\bibnamefont {Zhang}},\ }\href {\doibase https://doi.org/10.1088/1475-7516/2024/10/032} {\bibfield  {journal} {\bibinfo  {journal} {JCAP}\ }\textbf {\bibinfo {volume} {10}},\ \bibinfo {pages} {032} (\bibinfo {year} {2024})}\BibitemShut {NoStop}%
\bibitem [{\citenamefont {Jiang}\ \emph {et~al.}(2019)\citenamefont {Jiang}, \citenamefont {Wen},\ and\ \citenamefont {Chen}}]{Jiang2019}%
  \BibitemOpen
  \bibfield  {author} {\bibinfo {author} {\bibfnamefont {R.}~\bibnamefont {Jiang}}, \bibinfo {author} {\bibfnamefont {D.}~\bibnamefont {Wen}}, \ and\ \bibinfo {author} {\bibfnamefont {H.}~\bibnamefont {Chen}},\ }\href {\doibase 10.1103/PhysRevD.100.123010} {\bibfield  {journal} {\bibinfo  {journal} {Phys. Rev. D}\ }\textbf {\bibinfo {volume} {100}},\ \bibinfo {pages} {123010} (\bibinfo {year} {2019})}\BibitemShut {NoStop}%
\bibitem [{\citenamefont {Bagchi}(2011)}]{Bagchi2011}%
  \BibitemOpen
  \bibfield  {author} {\bibinfo {author} {\bibfnamefont {M.}~\bibnamefont {Bagchi}},\ }\href {\doibase 10.1111/j.1745-3933.2011.01030.x} {\bibfield  {journal} {\bibinfo  {journal} {MNRAS: Letters}\ }\textbf {\bibinfo {volume} {413}},\ \bibinfo {pages} {L47} (\bibinfo {year} {2011})}\BibitemShut {NoStop}%
\bibitem [{\citenamefont {Sotani}\ \emph {et~al.}(2011)\citenamefont {Sotani}, \citenamefont {Yasutake}, \citenamefont {Maruyama},\ and\ \citenamefont {Tatsumi}}]{Sotani2011}%
  \BibitemOpen
  \bibfield  {author} {\bibinfo {author} {\bibfnamefont {H.}~\bibnamefont {Sotani}}, \bibinfo {author} {\bibfnamefont {N.}~\bibnamefont {Yasutake}}, \bibinfo {author} {\bibfnamefont {T.}~\bibnamefont {Maruyama}}, \ and\ \bibinfo {author} {\bibfnamefont {T.}~\bibnamefont {Tatsumi}},\ }\href {\doibase 10.1103/PhysRevD.83.024014} {\bibfield  {journal} {\bibinfo  {journal} {Phys. Rev. D}\ }\textbf {\bibinfo {volume} {83}},\ \bibinfo {pages} {024014} (\bibinfo {year} {2011})}\BibitemShut {NoStop}%
\bibitem [{\citenamefont {Sotani}\ and\ \citenamefont {Takiwaki}(2020)}]{Sotani2020PRD}%
  \BibitemOpen
  \bibfield  {author} {\bibinfo {author} {\bibfnamefont {H.}~\bibnamefont {Sotani}}\ and\ \bibinfo {author} {\bibfnamefont {T.}~\bibnamefont {Takiwaki}},\ }\href {\doibase 10.1103/PhysRevD.102.063025} {\bibfield  {journal} {\bibinfo  {journal} {Phys. Rev. D}\ }\textbf {\bibinfo {volume} {102}},\ \bibinfo {pages} {063025} (\bibinfo {year} {2020})}\BibitemShut {NoStop}%
\bibitem [{\citenamefont {Kunjipurayil}\ \emph {et~al.}(2022)\citenamefont {Kunjipurayil}, \citenamefont {Zhao}, \citenamefont {Kumar}, \citenamefont {Agrawal},\ and\ \citenamefont {Prakash}}]{Kunjipurayil2022}%
  \BibitemOpen
  \bibfield  {author} {\bibinfo {author} {\bibfnamefont {A.}~\bibnamefont {Kunjipurayil}}, \bibinfo {author} {\bibfnamefont {T.}~\bibnamefont {Zhao}}, \bibinfo {author} {\bibfnamefont {B.}~\bibnamefont {Kumar}}, \bibinfo {author} {\bibfnamefont {B.~K.}\ \bibnamefont {Agrawal}}, \ and\ \bibinfo {author} {\bibfnamefont {M.}~\bibnamefont {Prakash}},\ }\href {\doibase 10.1103/PhysRevD.106.063005} {\bibfield  {journal} {\bibinfo  {journal} {Phys. Rev. D}\ }\textbf {\bibinfo {volume} {106}},\ \bibinfo {pages} {063005} (\bibinfo {year} {2022})}\BibitemShut {NoStop}%
\bibitem [{\citenamefont {Yagi}\ and\ \citenamefont {Yunes}(2017)}]{YagiYunes2017}%
  \BibitemOpen
  \bibfield  {author} {\bibinfo {author} {\bibfnamefont {K.}~\bibnamefont {Yagi}}\ and\ \bibinfo {author} {\bibfnamefont {N.}~\bibnamefont {Yunes}},\ }\href {\doibase https://doi.org/10.1016/j.physrep.2017.03.002} {\bibfield  {journal} {\bibinfo  {journal} {Phys. Rep.}\ }\textbf {\bibinfo {volume} {681}},\ \bibinfo {pages} {1} (\bibinfo {year} {2017})}\BibitemShut {NoStop}%
\bibitem [{\citenamefont {Abbott}\ \emph {et~al.}(2017)\citenamefont {Abbott} \emph {et~al.}}]{Abbott2017PRL}%
  \BibitemOpen
  \bibfield  {author} {\bibinfo {author} {\bibfnamefont {B.~P.}\ \bibnamefont {Abbott}} \emph {et~al.} (\bibinfo {collaboration} {LIGO Scientific Collaboration and Virgo Collaboration}),\ }\href {\doibase 10.1103/PhysRevLett.119.161101} {\bibfield  {journal} {\bibinfo  {journal} {Phys. Rev. Lett.}\ }\textbf {\bibinfo {volume} {119}},\ \bibinfo {pages} {161101} (\bibinfo {year} {2017})}\BibitemShut {NoStop}%
\bibitem [{\citenamefont {Chan}\ \emph {et~al.}(2016)\citenamefont {Chan}, \citenamefont {Chan},\ and\ \citenamefont {Leung}}]{chan2016universality}%
  \BibitemOpen
  \bibfield  {author} {\bibinfo {author} {\bibfnamefont {T.}~\bibnamefont {Chan}}, \bibinfo {author} {\bibfnamefont {A.~P.}\ \bibnamefont {Chan}}, \ and\ \bibinfo {author} {\bibfnamefont {P.}~\bibnamefont {Leung}},\ }\href {\doibase 10.1103/PhysRevD.93.024033} {\bibfield  {journal} {\bibinfo  {journal} {Phys. Rev. D}\ }\textbf {\bibinfo {volume} {93}},\ \bibinfo {pages} {024033} (\bibinfo {year} {2016})}\BibitemShut {NoStop}%
\bibitem [{\citenamefont {Breu}\ and\ \citenamefont {Rezzolla}(2016)}]{breu2016maximum}%
  \BibitemOpen
  \bibfield  {author} {\bibinfo {author} {\bibfnamefont {C.}~\bibnamefont {Breu}}\ and\ \bibinfo {author} {\bibfnamefont {L.}~\bibnamefont {Rezzolla}},\ }\href {\doibase 10.1093/mnras/stw575} {\bibfield  {journal} {\bibinfo  {journal} {MNRAS}\ }\textbf {\bibinfo {volume} {459}},\ \bibinfo {pages} {646} (\bibinfo {year} {2016})}\BibitemShut {NoStop}%
\bibitem [{\citenamefont {Ravenhall}\ and\ \citenamefont {Pethick}(1994)}]{ravenhall1994neutron}%
  \BibitemOpen
  \bibfield  {author} {\bibinfo {author} {\bibfnamefont {D.}~\bibnamefont {Ravenhall}}\ and\ \bibinfo {author} {\bibfnamefont {C.~J.}\ \bibnamefont {Pethick}},\ }\href {\doibase 1994ApJ...424..846R} {\bibfield  {journal} {\bibinfo  {journal} {Astrophys. J.}\ }\textbf {\bibinfo {volume} {424}},\ \bibinfo {pages} {846} (\bibinfo {year} {1994})}\BibitemShut {NoStop}%
\bibitem [{\citenamefont {Lattimer}\ and\ \citenamefont {Schutz}(2005)}]{lattimer2005constraining}%
  \BibitemOpen
  \bibfield  {author} {\bibinfo {author} {\bibfnamefont {J.~M.}\ \bibnamefont {Lattimer}}\ and\ \bibinfo {author} {\bibfnamefont {B.~F.}\ \bibnamefont {Schutz}},\ }\href {\doibase 10.1086/431543} {\bibfield  {journal} {\bibinfo  {journal} {Astrophys. J.}\ }\textbf {\bibinfo {volume} {629}},\ \bibinfo {pages} {979} (\bibinfo {year} {2005})}\BibitemShut {NoStop}%
\bibitem [{\citenamefont {Bejger}\ and\ \citenamefont {Haensel}(2002)}]{bejger2002moments}%
  \BibitemOpen
  \bibfield  {author} {\bibinfo {author} {\bibfnamefont {M.}~\bibnamefont {Bejger}}\ and\ \bibinfo {author} {\bibfnamefont {P.}~\bibnamefont {Haensel}},\ }\href {\doibase 10.1051/0004-6361:20021241} {\bibfield  {journal} {\bibinfo  {journal} {A\&A}\ }\textbf {\bibinfo {volume} {396}},\ \bibinfo {pages} {917} (\bibinfo {year} {2002})}\BibitemShut {NoStop}%
\bibitem [{\citenamefont {Staykov}\ \emph {et~al.}(2016)\citenamefont {Staykov}, \citenamefont {Doneva},\ and\ \citenamefont {Yazadjiev}}]{staykov2016moment}%
  \BibitemOpen
  \bibfield  {author} {\bibinfo {author} {\bibfnamefont {K.~V.}\ \bibnamefont {Staykov}}, \bibinfo {author} {\bibfnamefont {D.~D.}\ \bibnamefont {Doneva}}, \ and\ \bibinfo {author} {\bibfnamefont {S.~S.}\ \bibnamefont {Yazadjiev}},\ }\href {\doibase 10.1103/PhysRevD.93.084010} {\bibfield  {journal} {\bibinfo  {journal} {Phys. Rev. D}\ }\textbf {\bibinfo {volume} {93}},\ \bibinfo {pages} {084010} (\bibinfo {year} {2016})}\BibitemShut {NoStop}%
\bibitem [{\citenamefont {Popchev}\ \emph {et~al.}(2019)\citenamefont {Popchev}, \citenamefont {Staykov}, \citenamefont {Doneva},\ and\ \citenamefont {Yazadjiev}}]{popchev2019moment}%
  \BibitemOpen
  \bibfield  {author} {\bibinfo {author} {\bibfnamefont {D.}~\bibnamefont {Popchev}}, \bibinfo {author} {\bibfnamefont {K.~V.}\ \bibnamefont {Staykov}}, \bibinfo {author} {\bibfnamefont {D.~D.}\ \bibnamefont {Doneva}}, \ and\ \bibinfo {author} {\bibfnamefont {S.~S.}\ \bibnamefont {Yazadjiev}},\ }\href {\doibase 10.1140/epjc/s10052-019-6691-x} {\bibfield  {journal} {\bibinfo  {journal} {Eur. Phys. J. C}\ }\textbf {\bibinfo {volume} {79}},\ \bibinfo {pages} {1} (\bibinfo {year} {2019})}\BibitemShut {NoStop}%
\bibitem [{\citenamefont {Andersson}\ and\ \citenamefont {Kokkotas}(1998)}]{Andersson:1997rn}%
  \BibitemOpen
  \bibfield  {author} {\bibinfo {author} {\bibfnamefont {N.}~\bibnamefont {Andersson}}\ and\ \bibinfo {author} {\bibfnamefont {K.~D.}\ \bibnamefont {Kokkotas}},\ }\href {\doibase 10.1046/j.1365-8711.1998.01840.x} {\bibfield  {journal} {\bibinfo  {journal} {MNRAS}\ }\textbf {\bibinfo {volume} {299}},\ \bibinfo {pages} {1059} (\bibinfo {year} {1998})}\BibitemShut {NoStop}%
\bibitem [{\citenamefont {Lau}\ \emph {et~al.}(2010)\citenamefont {Lau}, \citenamefont {Leung},\ and\ \citenamefont {Lin}}]{Lau_2010}%
  \BibitemOpen
  \bibfield  {author} {\bibinfo {author} {\bibfnamefont {H.~K.}\ \bibnamefont {Lau}}, \bibinfo {author} {\bibfnamefont {P.~T.}\ \bibnamefont {Leung}}, \ and\ \bibinfo {author} {\bibfnamefont {L.~M.}\ \bibnamefont {Lin}},\ }\href {\doibase 10.1088/0004-637X/714/2/1234} {\bibfield  {journal} {\bibinfo  {journal} {Astrophys. J.}\ }\textbf {\bibinfo {volume} {714}},\ \bibinfo {pages} {1234} (\bibinfo {year} {2010})}\BibitemShut {NoStop}%
\bibitem [{\citenamefont {Chan}\ \emph {et~al.}(2014)\citenamefont {Chan}, \citenamefont {Sham}, \citenamefont {Leung},\ and\ \citenamefont {Lin}}]{chan2014multipolar}%
  \BibitemOpen
  \bibfield  {author} {\bibinfo {author} {\bibfnamefont {T.}~\bibnamefont {Chan}}, \bibinfo {author} {\bibfnamefont {Y.-H.}\ \bibnamefont {Sham}}, \bibinfo {author} {\bibfnamefont {P.}~\bibnamefont {Leung}}, \ and\ \bibinfo {author} {\bibfnamefont {L.-M.}\ \bibnamefont {Lin}},\ }\href {\doibase 10.1103/PhysRevD.90.124023} {\bibfield  {journal} {\bibinfo  {journal} {Phys. Rev. D}\ }\textbf {\bibinfo {volume} {90}},\ \bibinfo {pages} {124023} (\bibinfo {year} {2014})}\BibitemShut {NoStop}%
\bibitem [{\citenamefont {Yagi}\ and\ \citenamefont {Yunes}(2013{\natexlab{b}})}]{Yagi2013Science}%
  \BibitemOpen
  \bibfield  {author} {\bibinfo {author} {\bibfnamefont {K.}~\bibnamefont {Yagi}}\ and\ \bibinfo {author} {\bibfnamefont {N.}~\bibnamefont {Yunes}},\ }\href {\doibase 10.1126/science.1236462} {\bibfield  {journal} {\bibinfo  {journal} {Science}\ }\textbf {\bibinfo {volume} {341}},\ \bibinfo {pages} {365} (\bibinfo {year} {2013}{\natexlab{b}})}\BibitemShut {NoStop}%
\bibitem [{\citenamefont {Wen}\ \emph {et~al.}(2019)\citenamefont {Wen}, \citenamefont {Li}, \citenamefont {Chen},\ and\ \citenamefont {Zhang}}]{Wen2019}%
  \BibitemOpen
  \bibfield  {author} {\bibinfo {author} {\bibfnamefont {D.~H.}\ \bibnamefont {Wen}}, \bibinfo {author} {\bibfnamefont {B.~A.}\ \bibnamefont {Li}}, \bibinfo {author} {\bibfnamefont {H.~Y.}\ \bibnamefont {Chen}}, \ and\ \bibinfo {author} {\bibfnamefont {N.~B.}\ \bibnamefont {Zhang}},\ }\href {\doibase 10.1103/PhysRevC.99.045806} {\bibfield  {journal} {\bibinfo  {journal} {Phys. Rev. C}\ }\textbf {\bibinfo {volume} {99}},\ \bibinfo {pages} {045806} (\bibinfo {year} {2019})}\BibitemShut {NoStop}%
\bibitem [{\citenamefont {Abbott}\ \emph {et~al.}(2018)\citenamefont {Abbott} \emph {et~al.}}]{Abbott2018PRL}%
  \BibitemOpen
  \bibfield  {author} {\bibinfo {author} {\bibfnamefont {B.~P.}\ \bibnamefont {Abbott}} \emph {et~al.} (\bibinfo {collaboration} {The LIGO and Virgo Scientific Collaborations}),\ }\href {\doibase 10.1103/PhysRevLett.121.161101} {\bibfield  {journal} {\bibinfo  {journal} {Phys. Rev. Lett.}\ }\textbf {\bibinfo {volume} {121}},\ \bibinfo {pages} {161101} (\bibinfo {year} {2018})}\BibitemShut {NoStop}%
\bibitem [{\citenamefont {Zhang}\ and\ \citenamefont {Mann}(2021)}]{Zhang2021}%
  \BibitemOpen
  \bibfield  {author} {\bibinfo {author} {\bibfnamefont {C.}~\bibnamefont {Zhang}}\ and\ \bibinfo {author} {\bibfnamefont {R.~B.}\ \bibnamefont {Mann}},\ }\href {\doibase 10.1103/PhysRevD.103.063018} {\bibfield  {journal} {\bibinfo  {journal} {Phys. Rev. D}\ }\textbf {\bibinfo {volume} {103}},\ \bibinfo {pages} {063018} (\bibinfo {year} {2021})}\BibitemShut {NoStop}%
\end{thebibliography}
\end{document}